%% file: lsd.tex
\documentclass[aps, prx, twocolumn, reprint, amsmath, amssymb, superscriptaddress]{revtex4-2}

\usepackage{amsmath, amssymb}
\usepackage{braket}
\usepackage{bm}
\usepackage{amsthm}
\usepackage{todonotes}
\usepackage{ulem}
\usepackage{mathtools}

\newcommand{\invis}[1]{}  %
\newcommand{\killblue}[1]{#1}  %
\usepackage{subfig}

\theoremstyle{plain}

\newtheorem*{thm*}{Theorem}

\newtheorem*{lem*}{Lemma}

\newtheorem*{cor*}{Corollary}
\theoremstyle{remark}

\newtheorem*{rem*}{Remark}

\newcommand{\blue}[1]{\textcolor{blue}{#1}}

\begin{document}

\title[Efficient Quantum Circuits for Electronic Hamiltonian Simulation
without Pauli Expansion
]{Efficient Quantum Circuits for Electronic Hamiltonian Simulation \\
without Pauli Expansion
}

\author{Tamiya Onodera}
\email[]{tamiya.onodera@riken.jp}
\affiliation{RIKEN, Center for Computational Science, 2-1, Hirosawa, Wako, Saitama 351-0198, Japan}

\author{Takeshi Sato}
\affiliation{Department of Nuclear Engineering and Management, Graduate School of Engineering, The University of Tokyo, 7-3-1 Hongo, Bunkyo-ku, Tokyo 113-8656, Japan}
\affiliation{Photon Science Center, Graduate School of Engineering, The University of Tokyo, 7-3-1 Hongo, Bunkyo-ku, Tokyo 113-8656, Japan}
\affiliation{Research Institute for Photon Science and Laser Technology, The University of Tokyo, 7-3-1 Hongo, Bunkyo-ku, Tokyo 113-0033, Japan}
\affiliation{RIKEN, TRIP Headquarters, 2-1, Hirosawa, Wako, Saitama 351-0198, Japan}

\begin{abstract}
\input{lsd_abstract_ver3to}
\end{abstract}

\maketitle

\section{Introduction}
\input{lsd_introduction_ver3to}

\section{Ladder-string-pair diagonalization}
\label{sec:lasp}

We consider a {\it ladder string} on $n$ qubits, which is a tensor product of $n$ ladder operators from $\{\sigma_{01}, \sigma_{10}\}$, where 
$\sigma_{01} := \ket{0} \! \bra{1}$ and $\sigma_{10} := \ket{1} \! \bra{0}$.
An $n$-length ladder string is nicely represented  as $\ket{x}\bra{\bar{x}}$
with an $n$-digit binary $x$.

What we are interested in is a sum of a ladder string and its conjugate, weighted by conjugate coefficients $c$ and $c^*$, such as
\begin{align}
L^{(x:n)} (c) = c  \ket{x}\bra{\bar{x}} + c^*\ket{\bar{x}}\bra{x},
\end{align}
which we call an  {\it n-digit Lasp (Ladder-string-pair) operator}.
We write the superscript as $(x:n)$ to make it clear that $x$ is an $n$-digit binary, but we may simply write $(x)$ when no confusion arises. 
Note that $L^{(\bar{x}:n)} (c^*)=L^{(x:n)} (c)$
and that
$L^{(x_1:n)} (c_1)$ and $L^{(x_2:n)} (c_2)$ are commutative when $x_1 \neq x_2 \land x_1 \neq \overline{x_2}$.

It is important to observe that the operator $L^{(x:n)}$ can be diagonalized as shown below, a fact first noted by Sato et al.~\cite{sato2024hamiltonian}.
Let us write  $c=\gamma e^{i\phi}$, where $\gamma,\phi\in\mathbb{R}$ and $\gamma$ is not restricted to be nonnegative.
With this representation,
\begin{align}
&L^{(x:n)} (\gamma e^{i\phi}) \nonumber \\
&=  \gamma \Big( 
\frac{e^{i\phi/2}\ket{x} +  e^{-i\phi/2}\ket{\bar{x}} } {\sqrt{2}}
\frac{e^{-i\phi/2}\bra{x} +  e^{i\phi/2}\bra{\bar{x}} }{\sqrt{2}} \nonumber \\
&~~~~~~-~~ \frac{e^{i\phi/2}\ket{x} -  e^{-i\phi/2}\ket{\bar{x}} }{\sqrt{2}}
\frac{e^{-i\phi/2}\bra{x} -  e^{i\phi/2}\bra{\bar{x}} }{\sqrt{2}} \Big) \nonumber \\ 
&= \gamma ~U^{(x:n)} (\phi)~ ( \ket{1}\bra{1}^{\otimes {(n-1)}} \otimes Z)~ U^{(x:n)} (\phi)^\dagger
\label{eq:diag_Lop}
\end{align}
where %
\begin{align*}
& U^{(x:n)} (\phi)\ket{1}^{\otimes {(n-1)}}\ket{0} 
= \frac{e^{i\phi/2}\ket{x} + e^{-i\phi/2}\ket{\bar{x}}}{\sqrt{2}}, \\
& U^{(x:n)} (\phi)\ket{1}^{\otimes {(n-1)}} \ket{1}
= \frac{e^{i\phi/2}\ket{x} -  e^{-i\phi/2}\ket{\bar{x}}}{\sqrt{2}}.
\end{align*}
We refer to this transformation 
as {\it Lasp diagonalization}. We can then represent
the time evolution operator of $L^{(x:n)}(\gamma e^{i\phi})$, namely $\exp~(-itL^{(x:n)}(\gamma e^{i\phi}))$, as
\begin{align}
U^{(x:n)}(\phi)~ CRZ_{0}^{1, \dots, n-1} (2 \gamma t)~ U^{(x:n)}(\phi)^\dagger,
\label{eq:exp_Lop}
\end{align}
where $CRZ_{k}^{\bm{b}}(\theta)$ is a $\theta$-radian rotation gate about the Z axis on qubit~$k$ controlled by all the qubits in $\bm{b}$. In what follows, we assume $t=1$ for simplicity.

The unitary $U^{(x:n)}(\phi)$ can be realized using a circuit for preparing the $n$-qubit GHZ state, denoted as $G^{(n)}$, as
\begin{align}
(\otimes_{k=0}^{n-1} X_k^{x_i})~
RZ_{0}(-\phi)~G^{(n)} ~(\otimes_{k=1}^{n-1} X_k)
\label{eq:U}
\end{align}
where $X_k$ is the X gate acting on qubit~$k$, $x=x_{n-1} x_{n-2} \dots x_0~(x_i \in \{0,1\})$~\footnote{As mentioned earlier, we assume $x_{n-1}=0$},
$RZ_k(\theta)$ denotes the Z rotation gate with angle $\theta$ on qubit~$k$,
and $G^{(n)}$ performs the following two transformations,
\begin{align*}
& G^{(n)} \ket{0}^{\otimes {(n-1)}} \ket{0}
= \frac{\ket{0}^{\otimes n} + \ket{1}^{\otimes n}}{\sqrt{2}}, \\
& G^{(n)} \ket{0}^{\otimes {(n-1)}} \ket{1}
= \frac{\ket{0}^{\otimes n} -  \ket{1}^{\otimes n}}{\sqrt{2}}.
\end{align*}

\begin{figure}[htbp]
  \centering
  \subfloat[slope-shaped]{\includegraphics[width=0.15\textwidth]{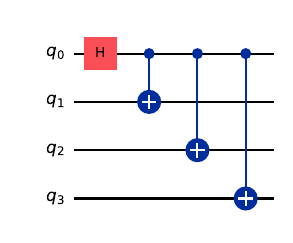}}
  \quad
  \subfloat[staircase-shaped]{\includegraphics[width=0.15\textwidth]{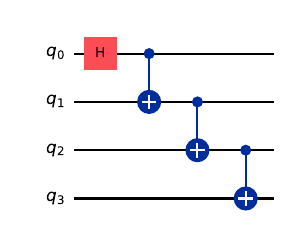}}  
  \subfloat[tree-shaped]{\includegraphics[width=0.125\textwidth]{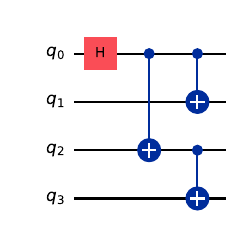}}   
  \caption{
  Three example circuits to prepare the 4-qubit GHZ state, 
  which we call slope-shaped, staircase-shaped, and tree-shaped from left to right, respectively. The Hadamard gate is in orange.
  }
  \label{fig:ghz}
\end{figure}

There are many different circuits to realize $G^{(n)}$. We show three of them for $G^{(4)}$ in Fig.~\ref{fig:ghz}, which we call, from left to right, slope-shaped, staircase-shaped, and tree-shaped, respectively. 
We note that, throughout the paper, Qiskit~\cite{qiskit2024}, an SDK for quantum computing, is used to construct and visualize the circuits in the figures.

\begin{figure}[htb]
\includegraphics[width=0.49\textwidth]{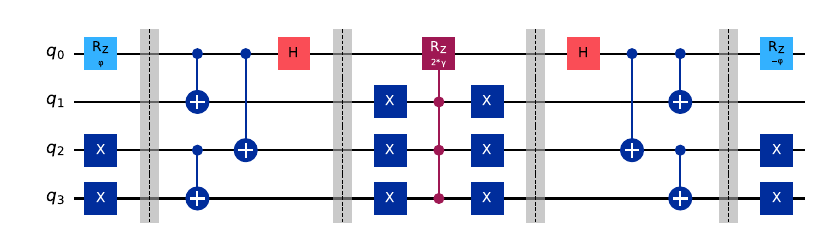}
\caption{
Time-evolution circuit for $L^{(1100)}(c)$, assuming that $c$ is represented  as $\gamma e^{i\phi}$ ($\gamma,\phi\in\mathbb{R}$). The tree-shaped circuit is chosen for the GHZ state preparation. The barriers are inserted to improve the visibility.
The X and RZ gates are in navy and light blue, respectively.
}
\label{fig:Lop1100}
\end{figure}

We present one of the many possible circuits implementing $\exp(-iL^{(1100)}(c))$ 
in Fig.~\ref{fig:Lop1100}, constructed faithfully following the above mentioned expressions with the tree-shaped circuit for $G^{(4)}$. Note that the barriers are inserted to improve visibility.

We note that there is, in fact, much greater flexibility 
in constructing the time-evolution circuit 
of $L^{(1100)}(c)$.
For instance, the RZ gate in Eq.~(\ref{eq:U})
can be placed on any qubit, not limited to qubit~0.
Furthermore,  we can choose an arbitrary qubit as the Z-rotation qubit in Eq.~(\ref{eq:exp_Lop}),
with $U^{(x:n)}$ and $G^{(n)}$ defined accordingly. 
We present the circuits with rotations on
qubit~1, qubit~2, and qubit~3
in Appendix~\ref{app:rot_on_diffq}.
Finally, we note that we can obtain the equivalent circuit by starting from $L^{(0011)}(c^*)$.

Yordanov et al.~\cite{Yordanov20} constructed a circuit with a similar structure
for what they call a double qubit excitation, the unitary evolution of an exponential of a parafermionic double excitation operator.
They constructed the circuit 
using an exchange-interaction circuit implementing
a single qubit excitation as a subcircuit.
Our Lasp-based formulation is much more general and can easily accommodate
this case as well,
as we describe in
\killblue{Appendix~\ref{app:yordanov}}.

Finally, we consider the CX count and depth of our general circuit when it is decomposed into single- and two-qubit gates. 
Since a C3RZ gate can be decomposed using eight CX gates (and eight Z-rotation gates), 
our circuit as shown in Fig~\ref{fig:Lop1100} can be decomposed into one with a CX count of 14 and a CX depth of 12, using the tree-shaped circuit for GHZ-state preparation.
Furthermore, we can apply the same circuit identity as that used by Yordanov et al.~\cite{Yordanov20} (specifically, the identity shown in Fig.~7 of Ref.~\cite{Yordanov20}), 
yielding a final reduction of one in both the CX count and depth.
Thus, our general Lasp-based construction achieves the same CX count and depth as the circuit of Yordanov et al.
Notice, however, that in this paper, we focus on high-level circuits in which the C3RZ gates are kept intact, rather than decomposed into single- and two-qubit gates. As we will see later, preserving these high-level structures is key to enabling significant circuit optimizations.

\section{Constructing circuits for electronic Hamiltonian simulation}
\label{sec:const_circ}

We describe a Lasp-based construction
of quantum circuits for electronic Hamiltonian simulation.
We consider the Hamiltonian in the form of second quantization given by
\begin{align*}
H = \sum_{p,q} h^p_q a_p^\dagger a_q +
\sum_{p,q,r,s} h^{pq}_{rs} a_p^\dagger a_q^\dagger a_r a_s
\end{align*}
where the annihilation $a_j$ and creation $a_j^\dagger$ operators obey the canonical anti-commutation relations \cite{helgaker2000molecular},
\begin{align}
\{ a_j, a_k^\dagger\} = \delta_{jk}I,~
\{ a_j, a_k\} =
\{ a_j^\dagger, a_k^\dagger\} =0.
\nonumber
\end{align}
We assume $n$ spin orbitals,
so the summation indices run over $[0,n-1]$.

To simulate the Hamiltonian on a quantum computer,
we construct a time-evolution circuit implementing $\exp(-iH)$.
In doing so, 
we first map a system of fermions to a system of qubits. 
As most commonly practiced, we use the Jordan-Wigner transform 
\cite{JordanWigner1928,Seeley2012}
and define $a_j$ and $a_j^\dagger$ as
\begin{align}
a_j = (\sigma_{01})_j \otimes_{k=0}^{j-1} Z_k,~~ %
a_j^\dagger = (\sigma_{10})_j \otimes_{k=0}^{j-1} Z_k
\nonumber
\end{align}
where we add the subscripts for the ladder operators to indicate which qubits they act on. These are then almost always subject to Pauli expansion \footnote{It substitutes the ladder operators as follows. 
\begin{align*}
\sigma_{01}=(X + iY)/2,~~~\sigma_{10}=(X - iY)/2
\end{align*}
}, followed by the concatenation of the time-evolution circuits for
the resulting Pauli strings. 

Rather than performing a Pauli expansion, we construct the circuits using Lasp diagonalization.
We start with the one-body Hamiltonian, mainly to illustrate our approach, and then proceed to the two-body Hamiltonian.
Throughout the paper, unless otherwise stated, we consider the most general case, where 
$h^p_q$ and $h^{pq}_{rs}$ are complex-valued.

\subsection{One-body Hamiltonian}

We consider the one-body Hamiltonian, 
\begin{align*}
\sum_{p,q} h^{p}_{q}~ a_p^\dagger a_q, 
\end{align*}
separately treating the cases of distinct and equal indices.

\subsubsection*{Distinct indices}

For mutually distinct $p$ and $q$, 
we transform the summation
as follows, 
distinguishing two cases of $p>q$ and $p<q$.
\begin{align*}
& \sum_{p,q} h^{p}_{q}~ a_p^\dagger a_q \\
=& \sum_{p>q} h^{p}_{q}~ a_p^\dagger a_q + 
\sum_{p<q} h^{p}_{q}~ a_p^\dagger a_q 
= \sum_{p>q} h^{p}_{q}~ a_p^\dagger a_q - 
\sum_{q>p} h^{p}_{q}~ a_q a_p^\dagger  \\
=& \sum_{d>e} (h^{d}_{e}~ a_d^\dagger a_e - 
 h^{e}_{d}~ a_d a_e^\dagger)   
= \sum_{p>q} (h^{p}_{q}~ a_p^\dagger a_q - 
h^{q}_{p}~ a_p a_q^\dagger) \\
\end{align*}
where we renamed the summation indices twice.
Since $h^q_p=(h^p_q)^*$ and $Z \sigma_{10}=-\sigma_{10}$,
the term in the summation can be written using a Lasp operator as
\begin{align*}
L^{(10)}_{p,q} (h^p_q) \otimes_{k=q+1}^{p-1} Z_k,
\end{align*}
where we add the subscripts for $L^{(10)}$ to indicate which qubits it acts on. 
We call the expression a {\it fermionic (two-digit) Lasp operator}, which we hereafter denote as $F^{(10)}_{p,q}(h^p_q)$.

Let us write  $h^p_q=\gamma^p_q \exp({i\phi^p_q})$, where $\gamma^p_q,\phi^p_q \in \mathbb{R}$.
Then, applying Eq.~(\ref{eq:diag_Lop}), we can diagonalize 
the fermionic Lasp operator to obtain
\begin{align}
\gamma^p_q~U&^{(10)}_{p,q} (\phi^p_q)~ \nonumber \\
&(\ket{1}\bra{1}_p \otimes_{k=q+1}^{p-1} Z_k \otimes Z_q)~ 
U^{(10)}_{p,q} (\phi^p_q)^\dagger,
\label{eq:diag_ferm_Lop}
\end{align}
where we add the subscripts for each $U^{(10)}$ 
to indicate which qubits it acts on. 
Note that, as usual, we suppress identity factors in tensor products throughout this paper.

Comparing to Eq.~(\ref{eq:diag_Lop}), we observe that the Z string of length $p-q-1$ is added in Eq.~(\ref{eq:diag_ferm_Lop}).  When constructing
the time-evolution circuit for the fermionic Lasp operator,
the effect of the Z string is handled using a standard technique in chemistry circuit construction, which  
encodes the parity of the qubits in the Z string into the target qubit of the Z rotation.
Let ${PE}_k^{\bm{b}}$ be a unitary 
to encode the parity of the qubits in $\bm{b}$ 
into qubit~$k$.  
We can then construct the time evolution circuit of the fermionic Lasp operator according to
\begin{align}
U^{(10)}_{p,q} & (\phi^p_q)~ (PE_q^{\mathcal{I}(q,p)})^\dagger \nonumber \\
&~CRZ_{q}^{p} (2\gamma^p_q) 
~ PE_q^{\mathcal{I}(q,p)}
~ U^{(10)}_{p,q}(\phi^p_q)^\dagger,
\label{eq:fLop10_diag}
\end{align}
where $\mathcal{I}(i,j):=
\{\, k \in \mathbb{Z} \mid i < k < j \,\}$.

\begin{figure}[htbp]
  \centering
  \subfloat[slope-shaped]{\includegraphics[width=0.15\textwidth]{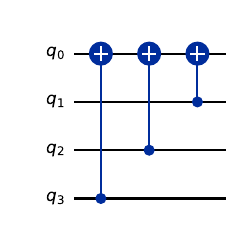}}
  \quad
  \subfloat[staircase-shaped]{\includegraphics[width=0.15\textwidth]{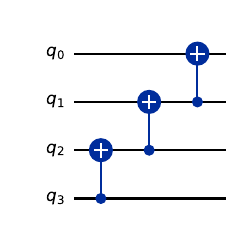}}  
  \subfloat[tree-shaped]{\includegraphics[width=0.125\textwidth]{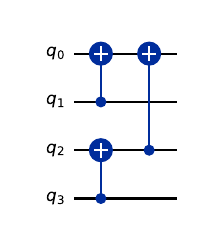}}   
  \caption{
  Three example circuits for $PE_0^{~1,2,3}$ which encodes the parity of $q_1$, $q_2$ and $q_3$ into $q_0$, which we call slope-shaped, staircase-shaped, and tree-shaped, from left to right, respectively. }
  \label{fig:pe}
\end{figure}

Similar to the GHZ-state preparation circuits, many different circuits can perform the parity encoding. We
show three example circuits for $PE_0^{1,2,3}$ in Fig.~\ref{fig:pe},
which we call,
from left to right, slope-shaped, staircase-shaped, and tree-shaped,
respectively. 

\begin{figure}[htbp]
  \centering
    \subfloat[A circuit for $\exp(-iF^{(10)}_{1,0}(c))$
    \label{fig:fop10_10}]{\includegraphics[width=0.36\textwidth]{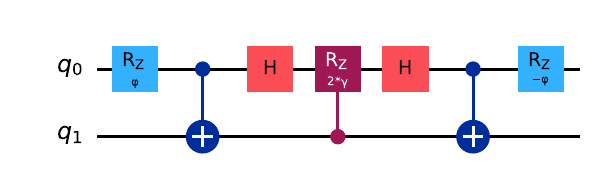}}
  \hfill
    \subfloat[A circuit for  $\exp(-iF^{(10)}_{4,1})(c))$ \label{fig:fLop10_41}]{\includegraphics[width=0.48\textwidth]{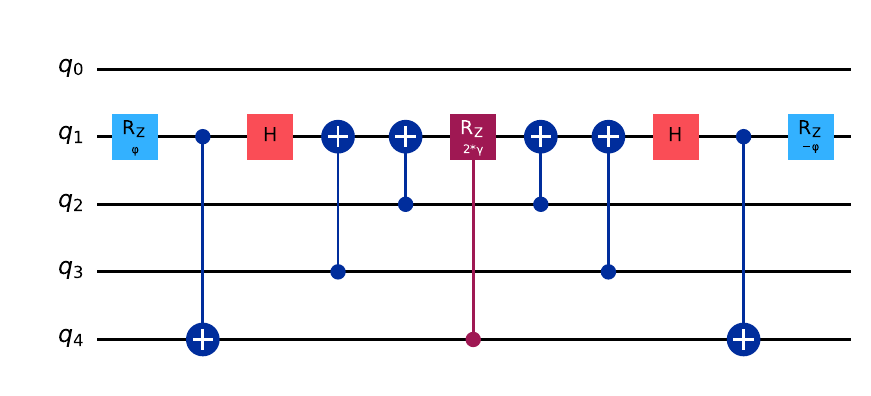}}
  \caption{
Time-evolution circuits for fermionic Lasp operators,
(a)~$F^{(10)}_{1,0}(c)$ and (b)~$F^{(10)}_{4,1}(c)$,
with $c$ represented as $\gamma e^{i\phi}$ $(\gamma,\phi\in\mathbb{R})$.
We do not need a circuit for the parity encoding in the former, while we use the slope-shaped circuit in the latter. We have canceled out the two X gates in $U^{(10)}$.  
These circuits correspond to
\killblue{four Pauli strings}
(with coefficients of two distinct absolute values)
in the Pauli expansion approach, and do not introduce any Trotter errors
at this stage.
}
  \label{fig:one}
\end{figure}

As examples, we show the time evolution circuits
for two fermionic Lasp operators,
$F^{(10)}_{1,0}(c)$ and $F^{(10)}_{4,1}(c)$, in Fig.~\ref{fig:one}.
Regarding the parity encoding,  we do not need it in the former, while we use the slope-shaped circuit in the latter. Note that we have canceled out the two X gates in $U^{(10)}$.

Since we consider the complex-valued case, these circuits correspond to
\killblue{four Pauli strings 
(with coefficients of two distinct absolute values)
in the Pauli-expansion approach.
Their nontrivial $X$ and $Y$ factors on qubits $p$ and $q$ are
$X \otimes X$, $X \otimes Y$, $Y \otimes X$, and $Y \otimes Y$,
while they all share the same Jordan--Wigner parity string
$\otimes_{k=q+1}^{p-1} Z_k$.}
Our circuits retain a higher-level structure and are more concise.
More importantly, when constructing a time-evolution circuit using the
Pauli-expansion approach, Trotter decomposition is generally used to
approximate the exponential of a sum of Pauli strings by a product of their
exponentials, introducing Trotter errors unless the relevant Pauli strings
commute. In contrast, our Lasp-based approach constructs a circuit equivalent
to the four Pauli strings without such an approximation and therefore
introduces no Trotter error at this stage.

\subsubsection*{Equal indices}

When $p=q$, we have the one-body Hamiltonian as
\begin{align*}
\sum_{p} h^{p}_{p}~ a_p^\dagger a_p
=
\sum_{p} h^{p}_{p}~ \ket{1}\bra{1}
\end{align*}
The time evolution circuit for the summation term can
simply be implemented by the phase gate, $P_p(-h^{p}_{p})$,
where $P_k(\theta)$ denotes the phase gate with angle $\theta$ on qubit~$k$.

\subsection{Two-body Hamiltonian} 

We now consider the two-body Hamiltonian:
\begin{align*}
& \sum_{p,q,r,s} \frac{h^{pq}_{rs}}{4}~ a_p^\dagger a_q^\dagger a_r a_s  
\end{align*}.
Since the summation term is zero when $p=q$ or $r=s$,
we can transform the summation as follows,
distinguishing between $p>q$ and $p<q$
and between $r>s$ and $r<s$,
\begin{align}
& \sum_{p,q,r,s} \frac{h^{pq}_{rs}}{4}~ a_p^\dagger a_q^\dagger a_r a_s  \nonumber \\
=& \sum_{p>q,~r>s} \frac{h^{pq}_{rs}}{4}~ a_p^\dagger a_q^\dagger a_r a_s + 
\sum_{p>q,~s>r} \frac{h^{pq}_{rs}}{4}~ a_p^\dagger a_q^\dagger a_r a_s \nonumber \\
+& \sum_{q>p,~r>s} \frac{h^{pq}_{rs}}{4}~ a_p^\dagger a_q^\dagger a_r a_s + 
\sum_{q>p,~s >r} \frac{h^{pq}_{rs}}{4}~ a_p^\dagger a_q^\dagger a_r a_s   \nonumber \\
=& \sum_{p>q,~r>s} \frac{h^{pq}_{rs}}{4}~ a_p^\dagger a_q^\dagger a_r a_s - 
\sum_{p>q,~s>r} \frac{h^{pq}_{rs}}{4}~ a_p^\dagger a_q^\dagger a_s a_r \nonumber \\
-& \sum_{q>p,~r>s} \frac{h^{pq}_{rs}}{4}~ a_q^\dagger a_p^\dagger a_r a_s +
\sum_{q>p,~s >r} \frac{h^{pq}_{rs}}{4}~ a_q^\dagger a_p^\dagger a_s a_r  \nonumber \\
=& \sum_{d>e,~f>g} ( \frac{h^{de}_{fg}}{4} -  \frac{h^{de}_{gf}}{4} -  \frac{h^{ed}_{fg}}{4} +  \frac{h^{ed}_{gf}}{4} )~ a_d^\dagger a_e^\dagger a_f a_g \nonumber \\
=& \sum_{p>q,~r>s} h^{pq}_{rs}~ a_p^\dagger a_q^\dagger a_r a_s  \label{eq:two_body}
\end{align}
where we renamed the summation indices twice 
and in the last equality we use
$h^{de}_{fg}=-  h^{de}_{gf} =-  h^{ed}_{fg} =  h^{ed}_{gf}$.

In the following, we consider three cases:
all four indices are distinct, exactly one pair of indices is equal,
and the indices form two equal pairs.

\subsubsection*{All Indices Distinct}

We further transform Eq. (\ref{eq:two_body}),
distinguishing cases according to the relative ordering of 
four indices,
\begin{align*}
& \sum_{p>q,~r>s} h^{pq}_{rs}~ a_p^\dagger a_q^\dagger a_r a_s  \\
=& \sum_{p>q > r>s} h^{pq}_{rs}~ a_p^\dagger a_q^\dagger a_r a_s + 
\sum_{p>r > q>s} h^{pq}_{rs}~ a_p^\dagger a_q^\dagger a_r a_s \\
&+
\sum_{p>r >s > q} h^{pq}_{rs}~ a_p^\dagger a_q^\dagger a_r a_s +
\sum_{r>p  > q >s} h^{pq}_{rs}~ a_p^\dagger a_q^\dagger a_r a_s \\
&+
\sum_{r>p >s > q} h^{pq}_{rs}~ a_p^\dagger a_q^\dagger a_r a_s +
\sum_{r>s >p > q} h^{pq}_{rs}~ a_p^\dagger a_q^\dagger a_r a_s    \\
=& \sum_{p>q > r>s} h^{pq}_{rs}~ a_p^\dagger a_q^\dagger a_r a_s  
- \sum_{p>r > q>s} h^{pq}_{rs}~ a_p^\dagger a_r a_q^\dagger  a_s \\
+& \sum_{p>r >s > q} h^{pq}_{rs}~ a_p^\dagger  a_r a_s a_q^\dagger 
+ \sum_{r>p  > q >s} h^{pq}_{rs}~ a_r a_p^\dagger a_q^\dagger a_s \\
-& \sum_{r>p >s > q} h^{pq}_{rs}~ a_r a_p^\dagger a_s a_q^\dagger  
+ \sum_{r>s >p > q} h^{pq}_{rs}~ a_r  a_s  a_p^\dagger a_q^\dagger \\
=& \sum_{d>e > f>g} (  h^{de}_{fg}~ a_d^\dagger a_e^\dagger a_f a_g  
-  h^{df}_{eg}~ a_d^\dagger a_e a_f^\dagger  a_g \\
&~~~~~~~~~~+   h^{dg}_{ef}~ a_d^\dagger  a_e a_f a_g^\dagger 
+  h^{ef}_{dg}~ a_d a_e^\dagger a_f^\dagger a_g \\
&~~~~~~~~~~ - h^{eg}_{df}~ a_d a_e^\dagger a_f a_g^\dagger  
+  h^{fg}_{de}~ a_d  a_e  a_f^\dagger a_g^\dagger )
\end{align*}
Since
$h^{de}_{fg}= (h^{fg}_{de})^*$,
$h^{df}_{eg}= (h^{eg}_{df})^*$ and
$h^{dg}_{ef}= (h^{ef}_{dg})^*$,
the expression in the summation 
can be represented using Lasp operators.
Noting $Z \sigma_{10}=-\sigma_{10}$ again and renaming the indices
$d,e,f$ and $g$ back to $p,q,r$ and $s$,
the expression is written as
\begin{align}
&L^{(1100)}_{p,q,r,s}(-h^{pq}_{rs})
\otimes_{k=q+1}^{p-1} Z_k \otimes_{k=s+1}^{r-1} Z_k \nonumber \\
+~ & L^{(1010)}_{p,q,r,s}(-h^{pr}_{qs})\otimes_{k=q+1}^{p-1} Z_k \otimes_{k=s+1}^{r-1} Z_k \label{eq:triad} \\
+~ & L^{(0110)}_{p,q,r,s}(-h^{qr}_{ps})\otimes_{k=q+1}^{p-1} Z_k \otimes_{k=s+1}^{r-1} Z_k. \nonumber
\end{align}
We call the summands fermionic (four-digit) Lasp operators,
denoting them as 
$F^{(1100)}_{p,q,r,s}(-h^{pq}_{rs})$,
$F^{(1010)}_{p,q,r,s}(-h^{pr}_{qs})$ and
$F^{(0110)}_{p,q,r,s}(-h^{qr}_{ps})$, respectively. 
 
\begin{figure}[htbp]
    \includegraphics[width=0.48\textwidth]{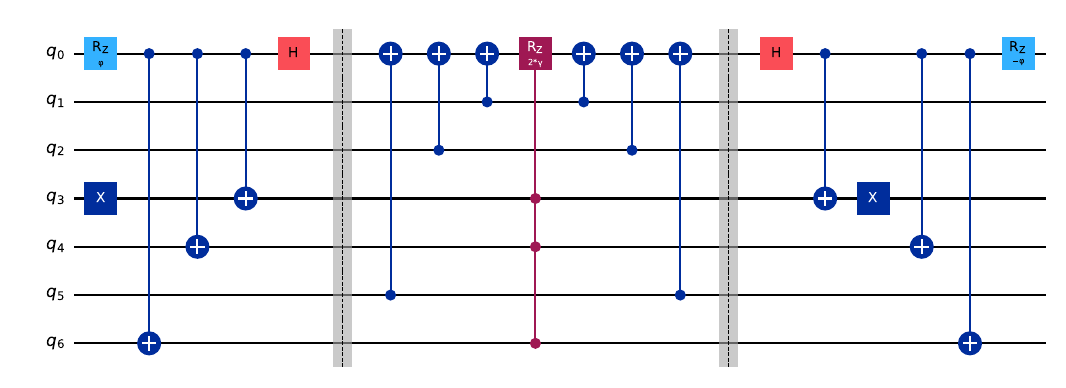}
  \caption{Time evolution circuit
  for $F^{(1100)}_{6,4,3,0}(c)$, 
  with $c$ represented as $\gamma e^{i\phi}$ 
$(\gamma,\phi\in\mathbb{R})$.
We adopt the slope-shaped circuits
for the GHZ state preparation and the parity encoding.  We optimized away as many X gates in $U^{(1100)}$. The barriers are inserted for better visibility.
The circuit corresponds to
\killblue{16 Pauli strings} 
(with coefficients of two distinct absolute values)
in the Pauli-expansion approach,
and does not introduce any Trotter errors at this stage.
}
\label{fig:fLop1100_6430}
\end{figure}

Using  the same procedure as for a fermionic two-digit Lasp operator,
we can construct the time-evolution circuit for 
$F^{(x)}_{p,q,r,s}(\gamma e^{i\phi})$ 
$(\gamma,\phi\in\mathbb{R})$
according to 
\begin{align}
U^{(x)}_{p,q,r,s}&(\phi)~ 
(PE_s^{\mathcal{I}(s,r) \cup \mathcal{I}(q,p)})^\dagger \nonumber \\
&~CRZ_{s}^{r,q,p} (2\gamma) 
~ PE_s^{\mathcal{I}(s,r) \cup \mathcal{I}(q,p)}
~ U^{(x)}_{p,q,r,s}(\phi)^\dagger. \nonumber
\end{align}
For example, let us construct the time evolution circuits for $F^{(1100)}_{3,2,1,0}(c)$ and $F^{(1100)}_{6,4,3,0}(c)$.
While Fig.~\ref{fig:Lop1100}, in fact, 
shows one possible circuit for the former,
Fig.~\ref{fig:fLop1100_6430} presents one for the latter,
where we adopt the slope-shaped circuits for $G^{(4)}$ and $PE^{1,2,5}_0$ and 
optimize away as many X gates in $U^{(1100)}$.
The barriers are inserted for improving the visibility.

Since we consider the complex-valued case, the time evolution circuit for
$F^{(x)}_{p,q,r,s}$
corresponds to
\killblue{16 Pauli strings}
(with coefficients of two distinct absolute values)
in the Pauli-expansion approach.
Our circuit retains a higher-level structure and is considerably more concise.
Furthermore,
no Trotter error
is introduced in
the Lasp-based circuit
even at this stage.

The entangling-gate cost of the time-evolution circuit for
$F^{(x)}_{p,q,r,s}$ comprises $2f(p,q,r,s)$ CX gates plus one C3RZ gate, where
$f(p,q,r,s)=(p-q+r-s-2)+3$. Finally,
our Lasp-based formulation naturally leads us to treat the three operators
in Eq.~(\ref{eq:triad}) collectively.
This results in a very efficient time-evolution circuit,
which we describe in the next section.

\subsubsection*{Exactly one pair equal}

We transform Eq. (\ref{eq:two_body}) in a similar manner, considering
all possible index configurations with exactly one equal pair.
\begin{align*}
& \sum_{p>q,~r>s} h^{pq}_{rs}~ a_p^\dagger a_q^\dagger a_r a_s  \\
=& \sum_{p=r>q>s} h^{pq}_{rs}~ a_p^\dagger a_q^\dagger a_r a_s + 
\sum_{p=r >s>q} h^{pq}_{rs}~ a_p^\dagger a_q^\dagger a_r a_s \\
&+
\sum_{r >p=s > q} h^{pq}_{rs}~ a_p^\dagger a_q^\dagger a_r a_s +
\sum_{p  > q=r >s} h^{pq}_{rs}~ a_p^\dagger a_q^\dagger a_r a_s \\
&+
\sum_{p >r > q=s} h^{pq}_{rs}~ a_p^\dagger a_q^\dagger a_r a_s +
\sum_{r>p > q=s} h^{pq}_{rs}~ a_p^\dagger a_q^\dagger a_r a_s    \\
=& -\sum_{p=r>q>s} h^{pq}_{rs}~ a_p^\dagger a_r a_q^\dagger  a_s + 
\sum_{p=r >s>q} h^{pq}_{rs}~ a_p^\dagger  a_r  a_s  a_q^\dagger\\
&-
\sum_{r >p=s > q} h^{pq}_{rs}~ a_r  a_p^\dagger  a_s a_q^\dagger+
\sum_{p  > q=r >s} h^{pq}_{rs}~ a_p^\dagger a_q^\dagger a_r a_s \\
&-
\sum_{p >r > q=s} h^{pq}_{rs}~ a_p^\dagger a_r a_q^\dagger a_s +
\sum_{r>p > q=s} h^{pq}_{rs}~ a_r a_p^\dagger a_q^\dagger a_s    \\
=& \sum_{d>e > f} (  - h^{de}_{df}~ a_d^\dagger a_d a_e^\dagger a_f  
+ h^{df}_{de}~ a_d^\dagger a_d a_e a_f^\dagger   \\
&~~~~~~~~~~ -  h^{ef}_{de}~ a_d  a_e^\dagger a_e a_f^\dagger 
+  h^{de}_{ef}~ a_d^\dagger a_e^\dagger a_e a_f \\
&~~~~~~~~~~ - h^{df}_{ef}~ a_d^\dagger a_e a_f^\dagger a_f  
+  h^{ef}_{df}~ a_d  a_e^\dagger  a_f^\dagger a_f )
\end{align*}

Since
$h^{de}_{df}= (h^{df}_{de})^*$,
$h^{ef}_{de}= (h^{de}_{ef})^*$, and
$h^{df}_{ef}= (h^{ef}_{df})^*$,
the expression in the summation 
can be represented using Lasp operators.
Noting $Z \sigma_{10}=-\sigma_{10}$ again and renaming the indices
$d,e,$ and $f$ back to $p,q,$ and $r$,
the expression is written as

\begin{align}
& \ket{1}\bra{1}_p \otimes L^{(10)}_{q,r}(-h^{pq}_{pr})
\otimes_{k=r+1}^{q-1} Z_k \nonumber \\
+~ &  \ket{1}\bra{1}_q \otimes L^{(10)}_{p,r}(-h^{pq}_{qr}) 
\otimes_{k=r+1}^{p-1} Z_k  \label{eq:triad2} \\
+~ &  \ket{1}\bra{1}_r \otimes  L^{(10)}_{p,q}(-h^{pr}_{qr})
\otimes_{k=q+1}^{p-1} Z_k . \nonumber
\end{align}

We see that each term is a
tensor product of $\ket{1}\bra{1}$
and
a two-digit fermionic Lasp operator.
When we consider the time evolution circuit,
the additional tensor factor of $\ket{1}\bra{1}$ 
simply adds another
control qubit for the Z-rotation gate.
Thus, for instance, we can construct the time-evolution circuit the third term according to 
\begin{align*}
U^{(10)}_{p,q} & (\phi^{pr}_{qr})~ (PE_q^{\mathcal{I}(q,p)})^\dagger \nonumber \\
&~CRZ_{q}^{r,p} (-2\gamma^{pr}_{qr}) 
~ PE_q^{\mathcal{I}(q,p)}
~ U^{(10)}_{p,q}(\phi^{pr}_{qr})^\dagger,
\end{align*}
where $h^{pr}_{qr}$ is represented  as 
$\gamma^{pr}_{qr} \exp({i\phi^{pr}_{qr})}$ 
$(\gamma^{pr}_{qr},\phi^{pr}_{qr} \in\mathbb{R})$.
The key difference from Eq.~(\ref{eq:fLop10_diag})
is that the Z-rotation gate is now controlled by
\killblue{qubits $p$ and $r$}.
The concrete time-evolution circuit for the third term with $(p,q,r)=(4,1,0)$
is the same as that in Fig.~\ref{fig:fLop10_41} except that the RZ gate is now controlled by qubits 4 and 0.

\subsubsection*{Two pairs equal}

The four indices $p>q$ and $r>s$ form two equal pairs
if and only if $p=r$ and $q=s$.
Thus, we have the two-body Hamiltonian as
\begin{align*}
&\sum_{p>q} h^{pq}_{pq}~ a_p^\dagger a_q^\dagger a_p a_q \\
=&
\sum_{p>q} -h^{pq}_{pq}~ \ket{1}\bra{1}_p \otimes\ket{1}\bra{1}_q
\end{align*}
The time evolution circuit for the summation term can be implemented by the controlled phase gate, 
$P_p(h^{pq}_{pq})$ controlled by qubit $q$ or 
$P_q(h^{pq}_{pq})$ controlled by qubit $p$.

\section{Optimizing circuits
for electronic Hamiltonian simulation }

The Lasp-based construction yields time-evolution circuits with high-level
structures that are thus very concise, whereas the Pauli-expansion-based construction
sometimes results in overwhelmingly lengthy circuits.
Recall that the circuit in Fig.~\ref{fig:fLop1100_6430} corresponds to
\killblue{16 Pauli strings} (with coefficients of two distinct absolute values).

Preserving high-level circuit structures allows us to expand the scope of optimization, thereby enabling more extensive optimizations that have been inaccessible with Pauli expansion. We demonstrate such optimizations in this section.

\subsection{Optimization for a triad}
\label{sec:triad}

\begin{figure*}[htbp]
  \centering
    \subfloat[Constructed by concatenating three individual circuits\label{fig:triad6430}]{\includegraphics[width=0.92\linewidth]{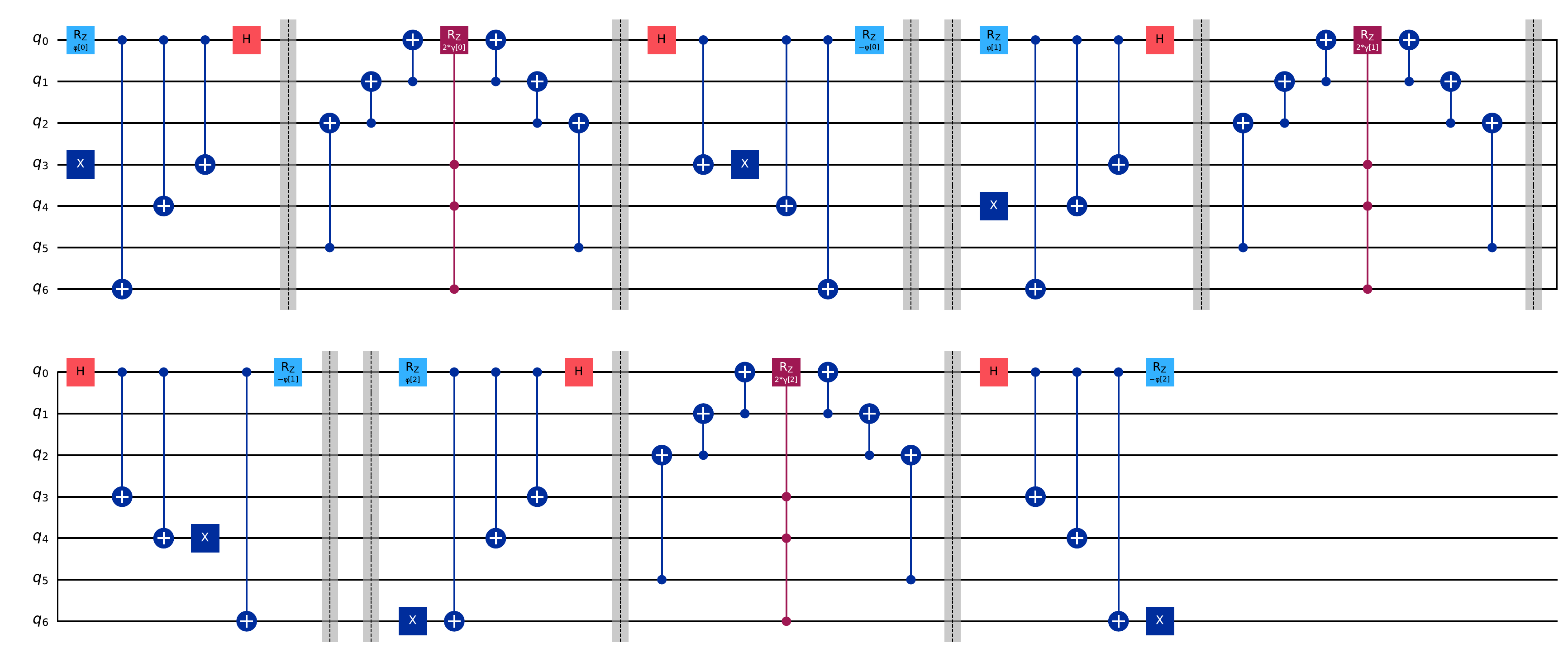}}
  \hfill $=$
    \subfloat[Optimized through gate commutation, gate cancellation, and local circuit rewriting\label{fig:triad6430_opt}]{\includegraphics[width=0.73\linewidth]{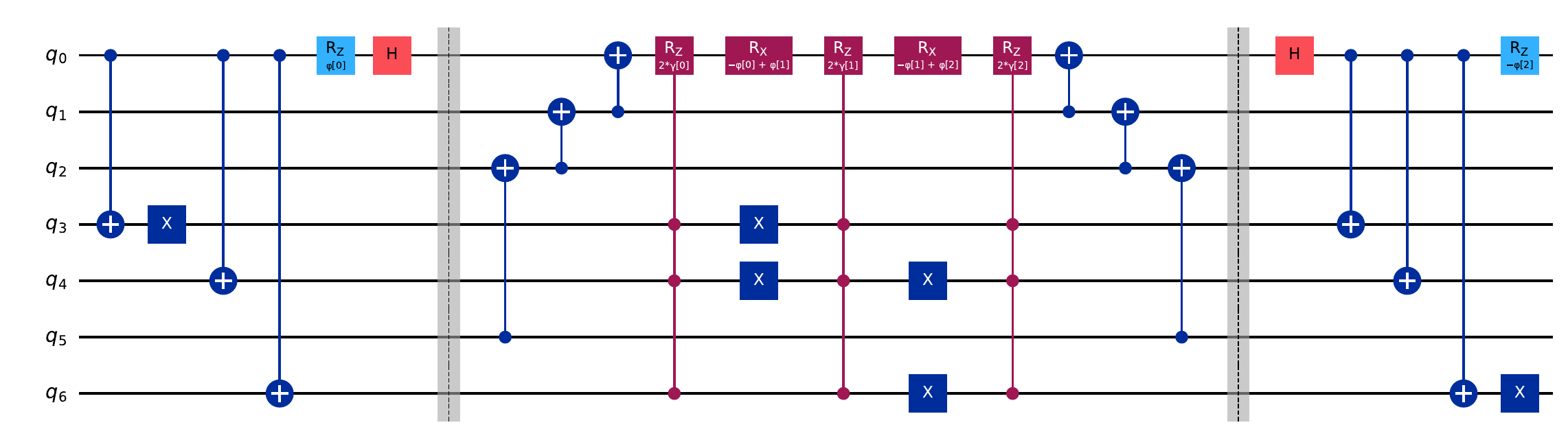}}
  \caption{
  Time-evolution circuits for a triad, or the sum of three fermionic Lasp operators, $F^{(1100)}(c[0])$,
$F^{(1010)}(c[1])$ and $F^{(0110)}(c[2])$, which share the same quadruple of summation indices, $(6,4,3,0)$.
We write $c[k]=\gamma[k] \exp ({i \phi[k])}$
$(\gamma[k],\phi[k] \in \mathbb{R})$
\killblue{for $k=0,1,2$}.
(a) Constructed by concatenating three individual circuits. 
(b) Optimized through gate commutation, gate cancellation, and local circuit rewriting.
Note that a CX gate commutes with a RZ gate on its control qubit.
All CX gates between adjacent C3RZ gates have been canceled, while one RX gate remains between each pair, with angles $-\phi[0]+\phi[1]$ and $-\phi[1]+\phi[2]$, respectively, in circuit order.
The CX gate count is reduced from 36 (above) to 12 (below),
while the C3RZ gate count remains three,
The barriers are inserted for better visibility.
Despite its concise form, the circuit corresponds to
\killblue{16 distinct Pauli strings}
(with coefficients of eight distinct absolute values)
after combining like terms in the Pauli-expansion approach.
In addition,
the circuit still does not suffer from Trotter errors since the three fermionic Lasp operators commute with each other.
}
\end{figure*}

We consider the time evolution circuit for
the sum of the three fermionic Lasp operators
in Eq.~(\ref{eq:triad}), which we refer to as a triad.
We first construct the circuit simply by concatenating the individual time-evolution circuits.
For instance, Fig.~\ref{fig:triad6430} 
shows the resulting circuit 
for $(p,q,r,s)=(6,4,3,0)$,
where we use the slope-typed for preparing the GHZ state and the staircase-typed for encoding the parity.
We clearly see many opportunities of CX-gate cancellation,
given that the a CX gate commutes a RZ gate acting on its control qubit.
In addition, after CX-gate cancellation, opportunities arise to apply local circuit rewrite rules to further reduce the CX-gate count. We defer the details of the optimization steps to Appendix~\ref{app:triad} and show the resulting optimized circuit in 
Fig.~\ref{fig:triad6430_opt}.
All CX gates between adjacent C3RZ gates
have been canceled. It is interesting to see that, if we used the slope-typed for the parity encoding, none of the CX gates for parity encoding could be eliminated.

Since we consider the complex-valued case, the time-evolution circuit
naively corresponds to
\killblue{48 Pauli strings}
in the Pauli-expansion approach, which reduce to
\killblue{16 distinct Pauli strings}
(with coefficients of eight distinct absolute values)
after combining like terms.
Notice that
our circuit retains a high-level structure and is very concise.
In addition,
while 
we deal with 
an exponential of the sum of 
the three fermionic Lasp operators,
these three operators commute
with each other, as we discussed
in Section~\ref{sec:lasp}.
Thus, the circuit for the triad
still does not suffer from Trotter errors.

Regarding the entangling-gate cost,
the time evolution circuit just concatenated
includes $6f(p,q,r,s)$ CX gates plus three C3RZ gates,
while the optimized circuit
includes $2f(p,q,r,s)$ CX gates plus three C3RZ gates.
Thus, by expanding the optimization scope
to the triad, 
we can significantly reduce the number of CX gates.

\subsection{Optimization for suitably ordered triads}
\label{sec:fleet}

We now expand the scope of our optimization to {\it suitably ordered} $O(n)$ triads.
For instance, denoting the triad in Eq~(\ref{eq:triad})
as $T_{p,q,r,s}$, we consider
\begin{align}
\prod_{k=p}^{n-1} \exp(- i T_{k,q,r,s}).
\label{eq:fleet}
\end{align}
\killblue{
Unlike the three operators within a single triad, distinct triads do not generally commute.
Eq.~(\ref{eq:fleet}) therefore represents a particular product-formula ordering of the triads.
Below, we show that this particular ordering enables substantial circuit optimization through cancellations across triads.
}

\begin{figure*}[htbp]
  \centering
    \subfloat[Constructed by concatenating time evolution circuits of four triads\label{fig:fleet}]{\includegraphics[width=0.96\linewidth]{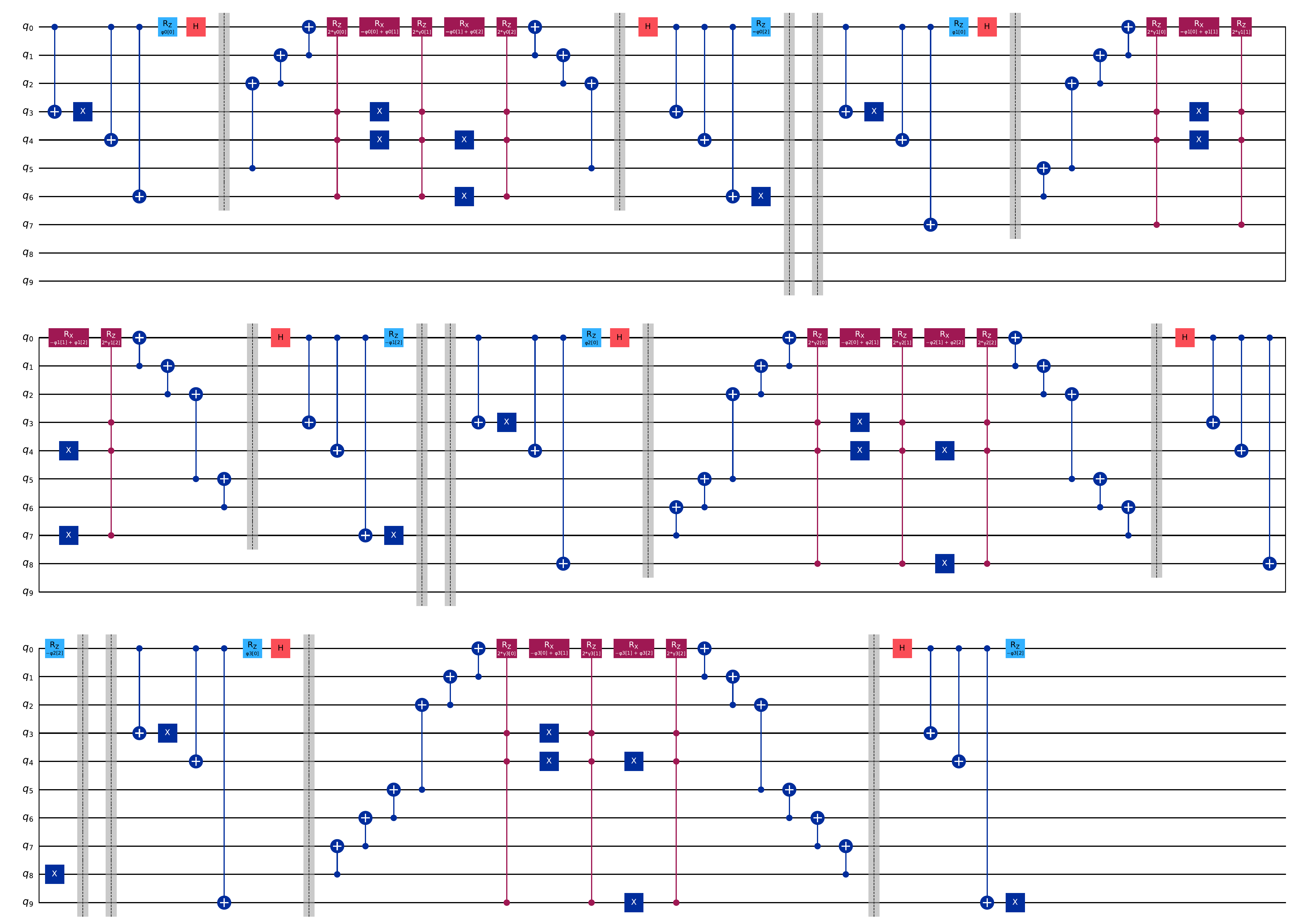}} \\ 
    \subfloat[Optimized through cascading CX reductions  \label{fig:fleet_opt}]{\includegraphics[width=0.96\linewidth]{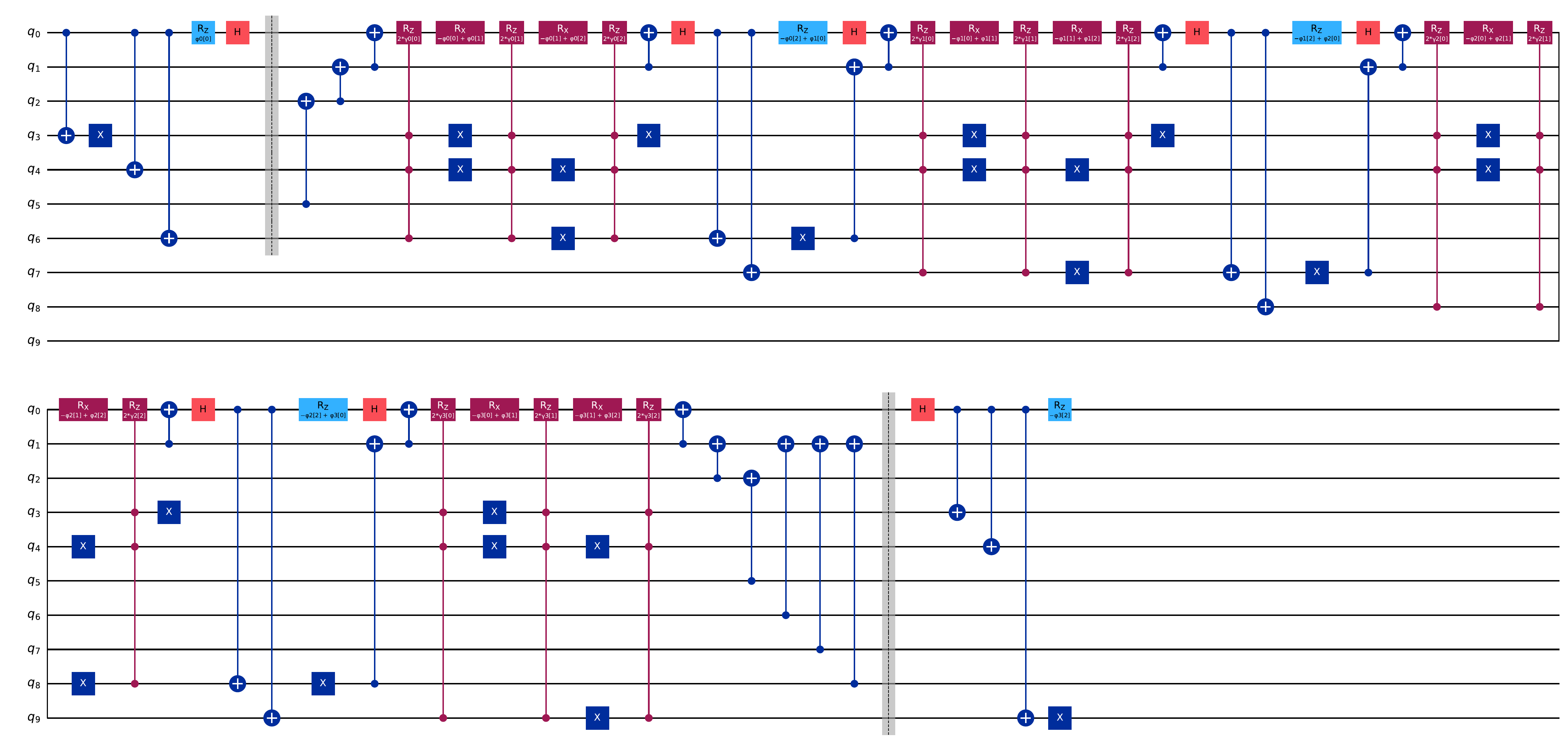}}
  \caption{
  Time evolution circuits for four triads, $\prod_{k=6}^{9} \exp(- i T_{k,4,3,0})$.
Three parameters of the $\bm{l}$-th triad $(\bm{l}=0,1,2,3)$ are 
$c\bm{l}[j]=\gamma\bm{l}[j] \exp({i \phi\bm{l}[j])}$ 
$(\gamma\bm{l}[j], \phi\bm{l}[j] \in\mathbb{R})$
for $j=0,1,2$.
(a) Constructed by concatenating time evolution circuits of four triads. The barriers are inserted for better visibility. 
(b) Optimized through cascading CX reductions. The CX gate count is reduced from 60 (above) to 30 (below).
}
\end{figure*}

If we just concatenate individual time evolution circuits of the triads (as optimized in the previous subsection), the resulting circuit ends up with $\sum_{k=p}^{n-1} 2f(k,q,r,s)$ CX gates
plus  $3(n-p)$ C3RZ gates. 
Assuming the eight-CX decomposition of a C3RZ gate, 
the circuit requires $O(n^2)$ CX gates 
when decomposed into single- and two-qubit gates.
For illustration, Fig.~\ref{fig:fleet}
shows the circuit thus concatenated 
for $n=10$ and $(p,q,r,s)=(6,4,3,0)$.

We now examine the boundary zone $B_k$
between the circuits for $\exp(- i T_{k,q,r,s})$
and $\exp(- i T_{k+1,q,r,s})$,
precisely
between the third $C3RZ$ gate of the former and the first $C3RZ$ gate of the latter.
$B_k$ composes the following operators (in the matrix multiplication order):
\begin{align}
&PE_s^{\mathcal{I}(s,r) \cup \mathcal{I}(q,k+1)} 
~U^{(1100)}_{k+1,q,r,s}(\phi^{k+1~q}_{rs})^\dagger \nonumber \\
&~~~~~~~~~U^{(0110)}_{k,q,r,s} (\phi^{qr}_{ks})~ (PE_s^{\mathcal{I}(s,r) \cup \mathcal{I}(q,k)})^\dagger, \nonumber 
\end{align}
where $h^{st}_{uv}$ is written  as 
$\gamma^{st}_{uv} \exp(i\phi^{st}_{uv})$,
with $\gamma^{st}_{uv},\phi^{st}_{uv} \in\mathbb{R}$,
for integers $s,t,u,v \ge 0$.
It is easy to see
that the six CX gates in the inner pair of $U^{(0110)}_{k,q,r,s}$ and 
$(U^{(1100)}_{k+1,q,r,s})^\dagger$
are reduced to two, $CX_{s,k}$ and $CX_{s, k+1}$, where $CX_{c,t}$ denotes a CX gate with 
control qubit $c$ and target qubit $t$.
We note that it is important to adopt the slope-shaped circuit for GHZ state preparation here.
The tree-shaped circuit would leave four CX gates,
while the staircase-shaped circuit leave six, with no CX gates canceled. 

Regarding 
the outer pair of $(PE_s^{\mathcal{I}(s,r) \cup \mathcal{I}(q,k)})^\dagger$ and 
$PE_s^{\mathcal{I}(s,r) \cup \mathcal{I}(q,k+1)}$,
it is surprising to see that the $2f(k,q,r,s)+1$ CX gates  
is eventually reduced to only three CX gates, $CX_{s+1,s}$, $CX_{k,s+1}$, and $CX_{s+1,s}$.
What happens is an intriguing cascade of CX-gate reductions, as explained in details in Appendix~\ref{app:cascade}.
Fig.~\ref{fig:fleet_opt}
shows the resulting optimized circuit.

The entangling gate cost of
the optimized circuit 
is $\{f(p,q,r,s)+f(n-1,q,r,s)+5(n-p-1)\}$
CX gates plus $3(n-p)$ C3RZ gates.
Thus, if decomposed into single- and two-qubit gates,
the time evolution circuit requires just $O(n)$ CX gates,
achieving the leading-order reduction in the CX gate count.

There will be many instances of suitably ordered triads.
For instance, for four small integers $0<s<t<u<v$, 
consider
\begin{align}
\prod_{k=0}^{n-v} \exp(- i T_{n-s,n-t,n-u,k}).
\label{eq:fleet2}
\end{align}
The corresponding circuit can then achieve similarly significant CX reductions
by choosing qubit $n-s$ as the Z-rotation qubit 
(Recall that any qubit can be chosen as the rotation qubit).
Furthermore, the circuits corresponding to Eqs.~(\ref{eq:fleet2}) and (\ref{eq:fleet})
roughly form upward- and downward-pointing triangles, respectively.
Thus, concatenating the two circuits in this order
is expected to yield a potentially significant reduction in circuit depth.

An instance of suitably ordered triads
should certainly have a fixed rotation qubit for all triads,
with adjacent triads differing in only one qubit,
whose index changes as little as possible.
The corresponding circuit is then expected to undergo significant CX reductions.
Despite these observations, a systematic investigation of suitable triad orderings and the composition of the resulting circuits is left for future work.

\subsection{Controlled Hamiltonian simulation}
\label{sec:cH}

Controlled versions of time-evolution circuits are frequently required as building blocks in quantum algorithms, notably in quantum phase estimation and interferometric measurements of correlation functions.
In general, a useful property of unitary conjugation is that,
for unitaries $P$ and $Q$, if $U=PQP^\dagger$,
the controlled version of $U$ can be constructed by adding a control
only to $Q$.
Thus, to construct a controlled version of the time-evolution circuit
for a fermionic Lasp operator shown in Fig.~\ref{fig:fLop1100_6430},
we only need to add a control to the C3RZ gate, making it the C4RZ gate.

To construct a controlled version of the time-evolution circuit for a triad,
we start with the circuit obtained by concatenating the individual circuits.
It is the same as Fig.~\ref{fig:triad6430}, except that the three $C3RZ$ gates
are now replaced by $C4RZ$ gates.
We can then apply exactly the same optimizations
that transform Fig.~\ref{fig:triad6430} into Fig.~\ref{fig:triad6430_opt},
resulting in the same circuit as in Fig.\ref{fig:triad6430_opt},
except that we now have three $C4RZ$ gates.

In this way, our Lasp-based approach
to time-evolution circuits for electronic Hamiltonian simulation
enables a simple and clear construction of their controlled versions.

\subsection{The Real-Valued Case}
\label{sec:real}

\begin{figure}[htbp]
\includegraphics[width=0.48\textwidth]{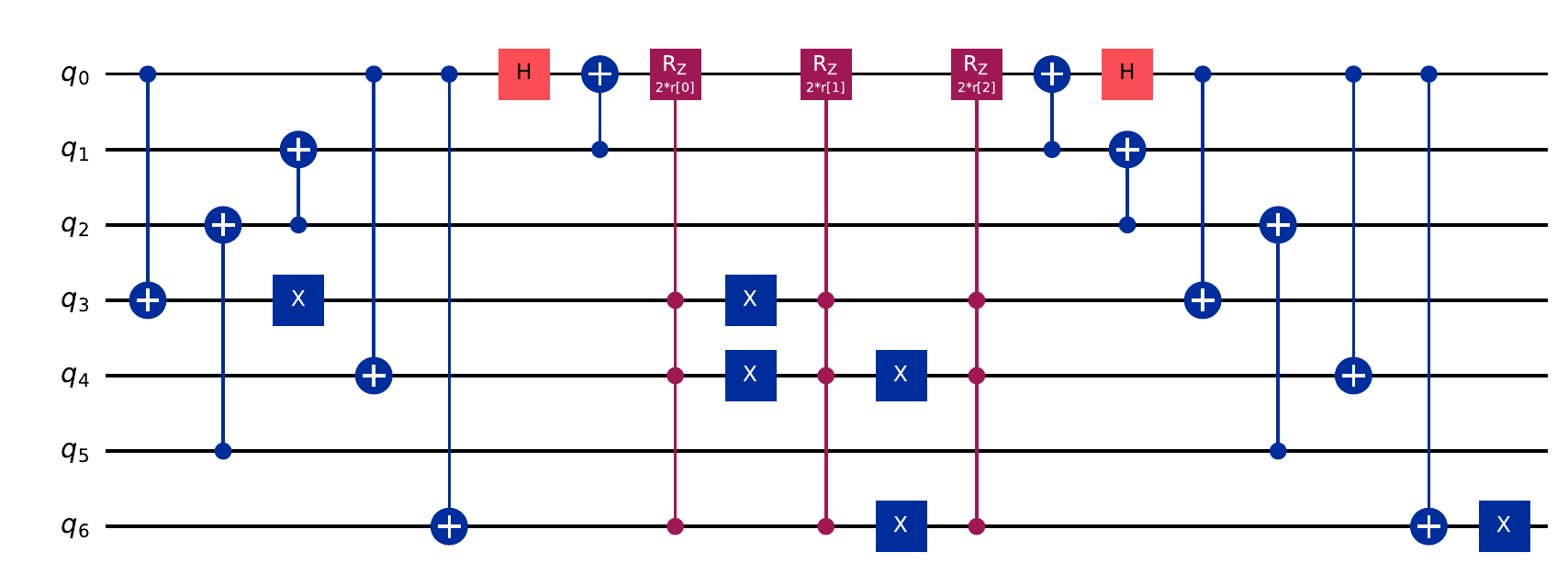}
\caption{
Time-evolution circuit for a triad in the real-valued case. 
The rotation angles of C3RZ gates are $r[j]$ $(j=0,1,2)$ from left to right.
The circuit corresponds to
\killblue{eight distinct Pauli strings}
(with coefficients of four distinct absolute values)
after combining like terms in the Pauli-expansion approach, and
does not suffer from Trotter errors.
}
\label{fig:triad6430_r_opt}
\end{figure}

While we focus on the complex-valued case throughout this paper, in this subsection we consider the real-valued case,
The time-evolution circuit 
for the real-valued case does not require RZ gates.
For $(p,q,r,s)=(6,4,3,0)$,
the circuit for a triad is shown
in Fig.~\ref{fig:triad6430_r_opt}
It corresponds to
\killblue{24 Pauli strings}
in the Pauli-expansion approach, which reduce to
\killblue{eight distinct Pauli strings}
(with coefficients of four distinct absolute values)
after combining like terms.
Like that in Fig.~\ref{fig:triad6430_opt},
the circuit does not suffer from Trotter errors.

At the high level, 
the entangling-gate cost of the real-valued circuit for a triad is the same as that of the complex-valued circuit, requiring $2f(p,q,r,s)$ CX gates plus three C3RZ gates.
When they are decomposed into single- and two-qubit gates, however, we see a significant difference. Note that the three C3RZ gates are now back-to-back in the real-valued case.
In addition,
these gates are mutually exclusive in the sense that at most one of the Z rotations is applied on the target qubit
for any computational basis state.
The three C3RZ gates therefore form an instance of a {\it uniformly controlled rotation}~\cite{Mottonen04},
and can be decomposed using only eight CX gates according to their recipe
based on the binary-reflected Gray code.
Thus, in the real-valued case, the time-evolution circuit for a triad can be realized with $\{2f(p,q,r,s)+8\}$ CX gates, along with a number of single-qubit gates.
Compared with the case where the eight-CX decomposition is applied to each C3RZ gate individually, 
the reduction is significant.
We explain how a uniformly controlled rotation is decomposed in
\killblue{Appendix~\ref{app:threeC3RZs}}.

\section{Summary}
\label{sec:summary}

\input{lsd_summary_ver3to}

\section*{Acknowledgment}
This work is supported in part by project JPNP20017, funded by the New Energy and Industrial Technology Development Organization (NEDO).
\input{lsd_acknowledgment_sato}
We acknowledge the use of IBM Quantum service for this work. The views expressed are those of the authors, and do not reflect the official policy or position of IBM or the IBM Quantum team. 

\appendix

\section{
Time-evolution circuits of a Lasp operator
with rotations on different qubits}
\label{app:rot_on_diffq}

\begin{figure}[htbp]
  \centering
  \subfloat[rotation on $q1$]{\includegraphics[width=0.48\textwidth]{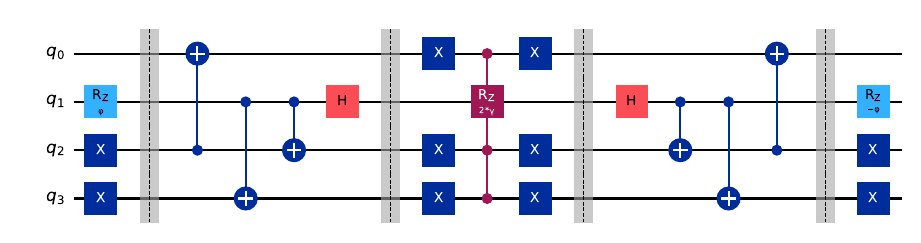}}
  \\
  \subfloat[rotation on $q2$]{\includegraphics[width=0.48\textwidth]{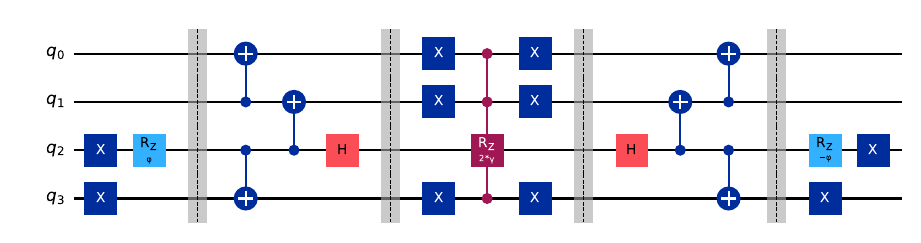}} 
  \\
  \subfloat[rotation on $q3$]{\includegraphics[width=0.48\textwidth]{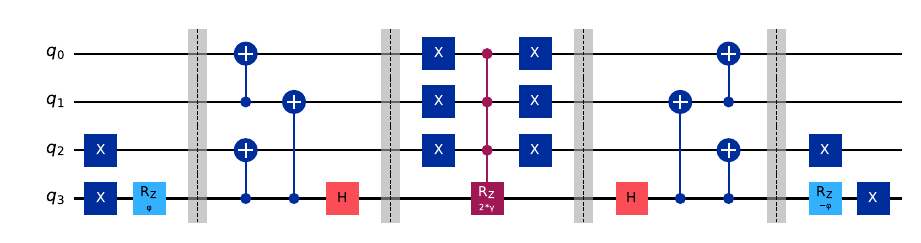}}   
  \caption{
  Time-evolution circuits of $L^{(1100)}(c)$ with rotations on different qubits, 
  with $c$ represented as $\gamma e^{i\phi}$ 
$(\gamma,\phi\in\mathbb{R})$.
   (a) $q0$, (b) $q1$, and (c) $q2$.  The barriers are inserted for better visibility. }
  \label{fig:Lop1100_qx}
\end{figure}

In constructing the time-evolution
circuit of a Lasp operator,
any qubit can be chosen
as the Z-rotation qubit
in Eq.~(\ref{eq:exp_Lop}),
with $U^{(x:n)}$ and $G^{(n)}$ defined accordingly.
Fig.~\ref{fig:Lop1100}
shows the time-evolution circuit
of $L^{(1100)}(c)$
with the rotation on $q_0$.
The other three circuits are shown
in Fig.~\ref{fig:Lop1100_qx},
which, from left to right, have the rotations
on $q_1$, $q_2$, and $q_3$, respectively.
All four circuits are equivalent.

\section{Lasp-based construction of a circuit for a double-qubit excitation}
\blue{\label{app:yordanov}}

Yordanov et al. \cite{Yordanov20} construct a circuit for the unitary evolution of an exponential of a $\theta$-parametrized parafermionic double excitation operator defined by a skew Hermitian operator (Fig.5 of \cite{Yordanov20}).  Precisely, they construct a circuit for $\exp(\tilde{T}^{pq}_{rs}(\theta))$ where $\tilde{T}^{pq}_{rs}$ is defined in our notation as follows.
\begin{align*}
\tilde{T}^{pq}_{rs}(\theta)
= \theta \ket{1100}\bra{0011}_{p, q, r, s} - 
\theta \ket{0011}\bra{1100}_{p, q, r, s} \\
\end{align*}
We then have
\[
\killblue{
\begin{aligned}
\exp(\tilde{T}^{pq}_{rs}(\theta))
&= \exp[-i(i\tilde{T}^{pq}_{rs}(\theta))] \\
&= \exp[-i L^{(1100)}_{p,q,r,s}(\theta e^{i \pi/2})].
\end{aligned}
}
\]
We can thus construct the circuit for $\exp(\tilde{T}^{pq}_{rs}(\theta))$ with our general
Lasp-based formulation, resulting in that in Fig.~\ref{fig:Lop1100}
with $\gamma=\theta$ and $\phi=\pi/2$.
It should be equivalent to the circuit in Fig. 5 of \cite{Yordanov20}.

\begin{figure*}[htbp]
  \includegraphics[width=0.76\textwidth]{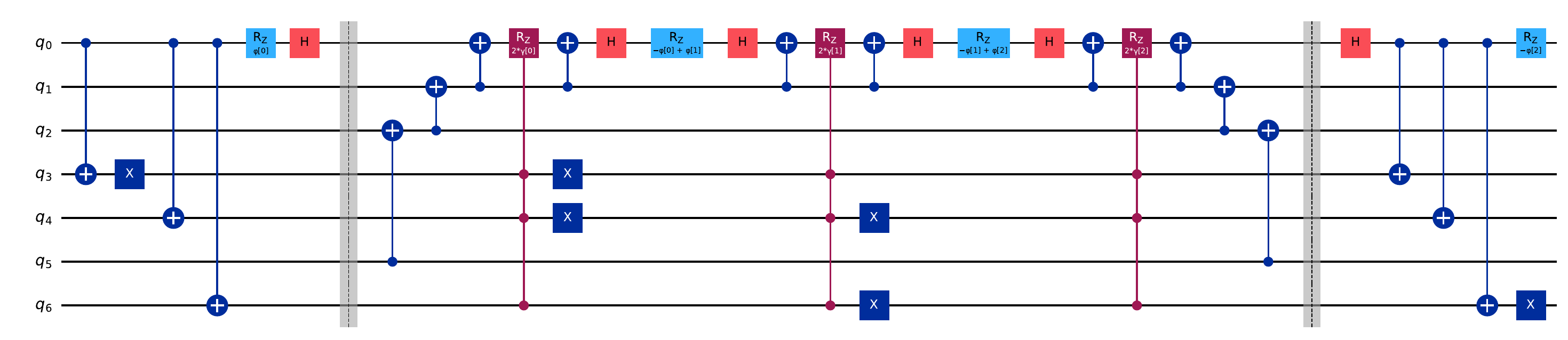}
  \caption{
  Time-evolution circuit for a triad
  after optimizing the circuit
  in Fig.~\ref{fig:triad6430}
  through gate commutation and cancellation.
}
\label{fig:triad6430_opt0}
\end{figure*}

\section{Optimization of a triad circuit}
\label{app:triad}

We explain how the time-evolution circuit for a triad is optimized from the circuit in Fig.~\ref{fig:triad6430}, obtained by concatenating three individual circuits, to the optimized circuit in Fig.~\ref{fig:triad6430_opt}. First, Fig.~\ref{fig:triad6430_opt0} shows the circuit after gate commutation and cancellation. We observe that the gate sequence $H_0RZ_0(\cdot)H_0$ occurs twice, where $H_k$ denotes the Hadamard gate on qubit $k$. Each occurrence can be rewritten as $RX_0(\cdot)$.
As a result, two instances of the gate sequence $CX_{1,0} RX_0(\cdot)  CX_{1,0}$ emerge. 
Since a CX gate commutes with a RX gate on its target qubit,
each sequence simplifies to just $RX_0(\cdot)$. This yields the optimized circuit shown in Fig.~\ref{fig:triad6430_opt}.

\section{Cascade of CX reductions
for suitably ordered triads}
\label{app:cascade}

In Section~\ref{sec:fleet}, we considered  the time evolution circuit of Eq.~(\ref{eq:fleet})
and discussed that the CX gate count of each boundary zone $B_k$
can be reduced to only five.
In this appendix, we explain what we mentioned is surprising, that is,
why the CX gate count for the parity encoding sub-circuits in $B_k$
is reduced to three.

\begin{figure}[htbp]
  \centering
    \subfloat[\label{fig:equiv_rel_a}]{\includegraphics[width=0.60\linewidth]{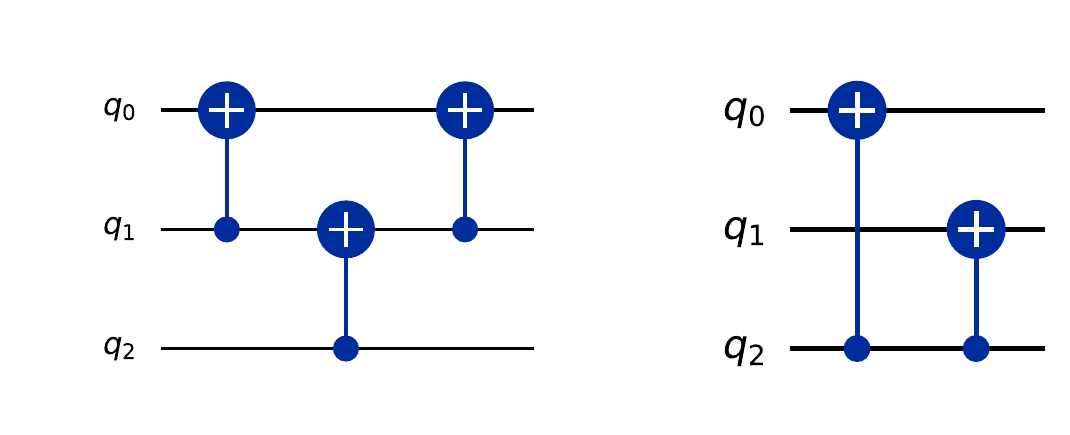}} \\
    \subfloat[\label{fig:equiv_rel_b}]{\includegraphics[width=0.60\linewidth]{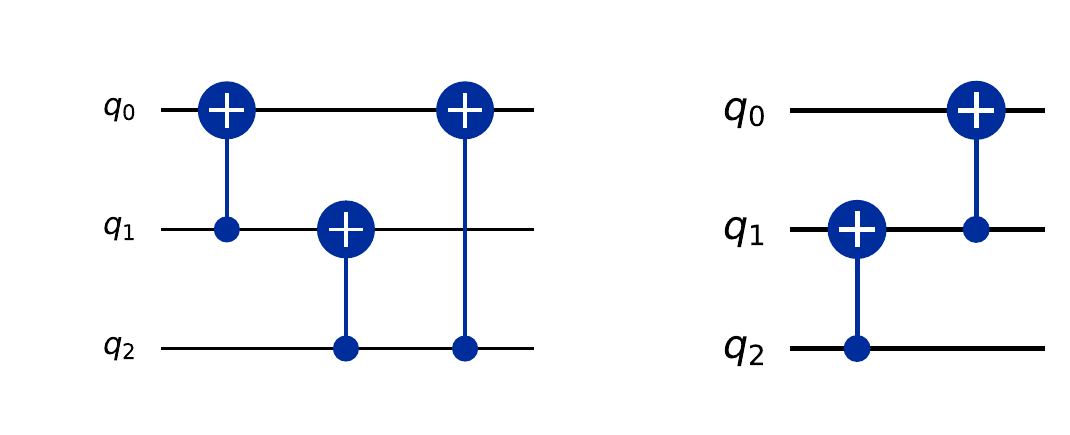}} \\
    \subfloat[\label{fig:equiv_rel_c}]{\includegraphics[width=0.60\linewidth]{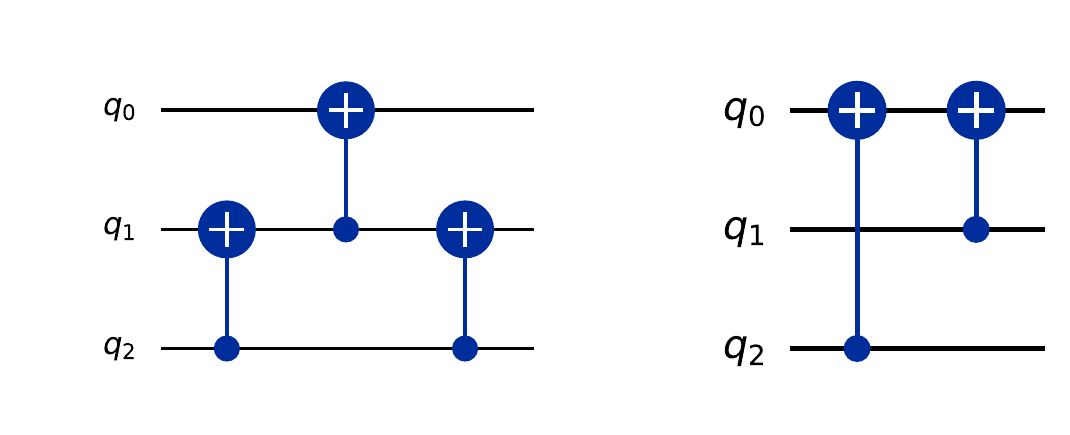}}
  \caption{
Equivalence relations between the left and right circuits.
In fact, each relation implies the other two.
}

\label{fig:equiv_rels}
\end{figure}

\begin{figure}[htbp]
  \centering
    \subfloat[\label{fig:CX_pat_a}]{\includegraphics[width=0.90\linewidth]{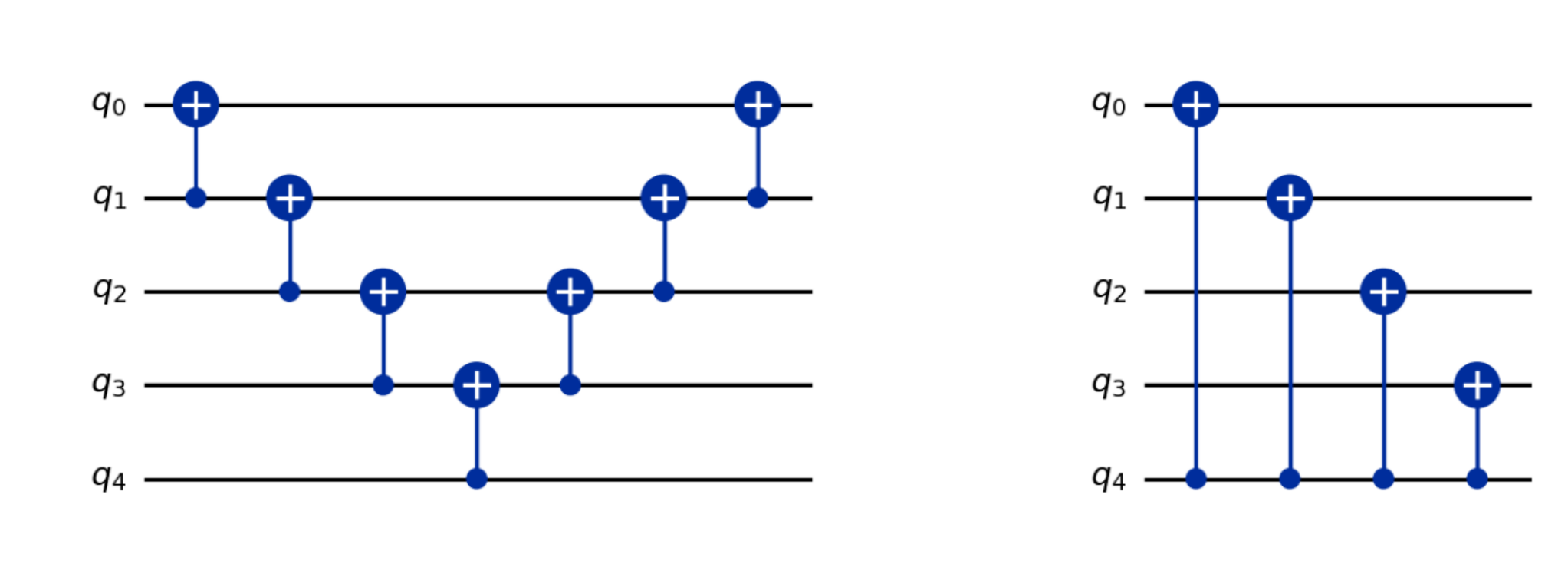}} \\
    \subfloat[\label{fig:CX_pat_b}]{\includegraphics[width=0.90\linewidth]{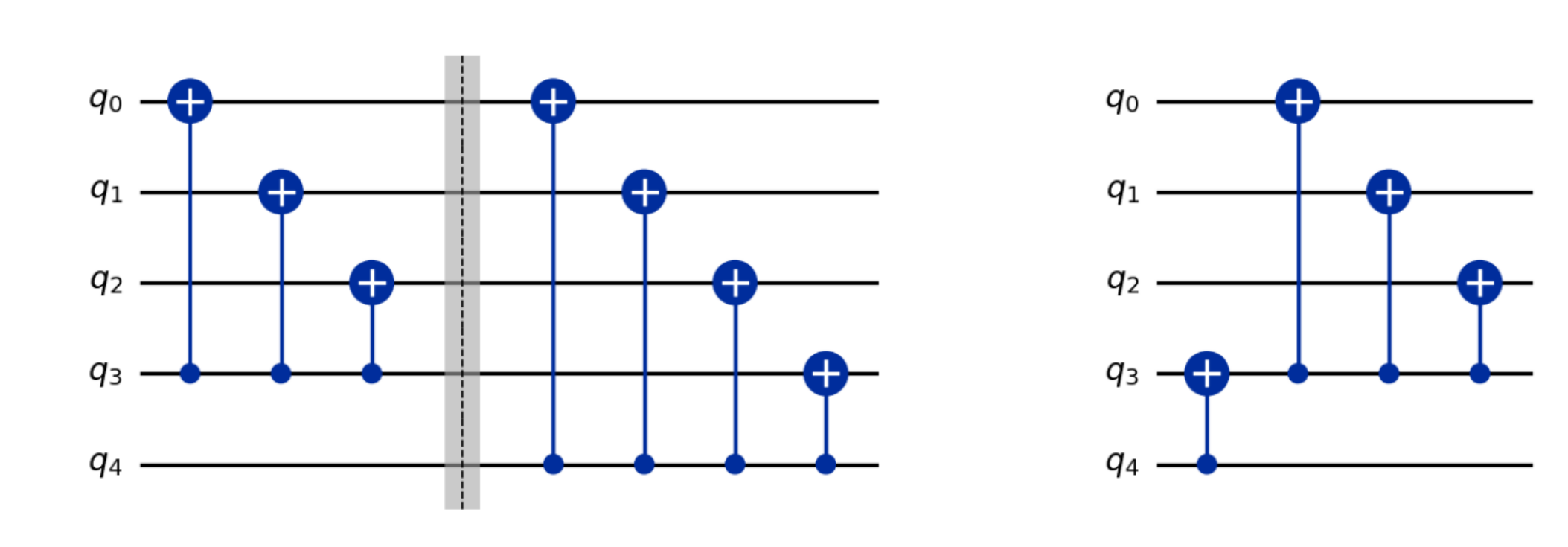}} \\
    \subfloat[\label{fig:CX_pat_c}]{\includegraphics[width=0.98\linewidth]{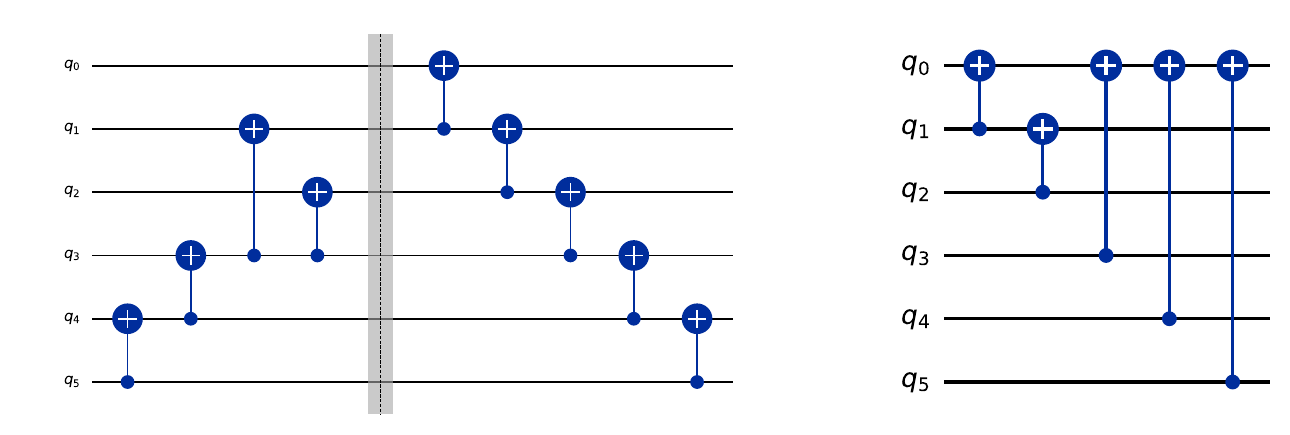}}
  \caption{
CX-pattern transforms from the left circuit to the right circuit.
The barriers are inserted for better visibility.
In (a), (b), and (c)
the right circuit can be obtained from the left circuit by repeatedly applying the equivalence relation
in Fig.~\ref{fig:equiv_rels}(a), (b), and (c),
respectively
}

\label{fig:CX_pats}
\end{figure}

We first show well-known equivalence relations in Fig.~\ref{fig:equiv_rels}.
Each relation actually implies the other two.
We then derive 
three CX-pattern transforms as in Fig.~\ref{fig:CX_pats}, using the relations.
In the first pattern transform,
the input is a staircase pattern that first descends and then ascends (hereafter a V-staircase pattern). 
By repeatedly applying the relation in Fig.~\ref{fig:equiv_rel_a}, 
we obtain a slope pattern of the same width
with nearly half as many CX gates. Here we call the number of qubits involved in a pattern the width of the pattern.
In the second pattern,
the input consists of two consecutive slopes,
with the second being one qubit wider downward than the first.
By repeatedly applying the relation in Fig.~\ref{fig:equiv_rel_b},
we obtain a pattern of a (one-step) ascending staircase and a slope, with nearly half as many CX gates.
Note that the CX gates in a slope commute,
so the slope can be ascending, descending, or arbitrarily ordered.
In the third pattern transform,
the input consists of two parts:
the first comprises an ascending staircase and a slope,
while the second is a descending staircase that is one qubit wider upward than the first.
By repeatedly applying the relation in Fig.~\ref{fig:equiv_rel_c},
we obtain a pattern of a descending staircase and a clothesline, with nearly half as many CX gates. 

\begin{figure*}[htbp]
  \centering
    \subfloat[After applying the pattern transform in Fig.~\ref{fig:CX_pat_a}\label{fig:fleet_opt_a}]{\includegraphics[width=0.98\linewidth]{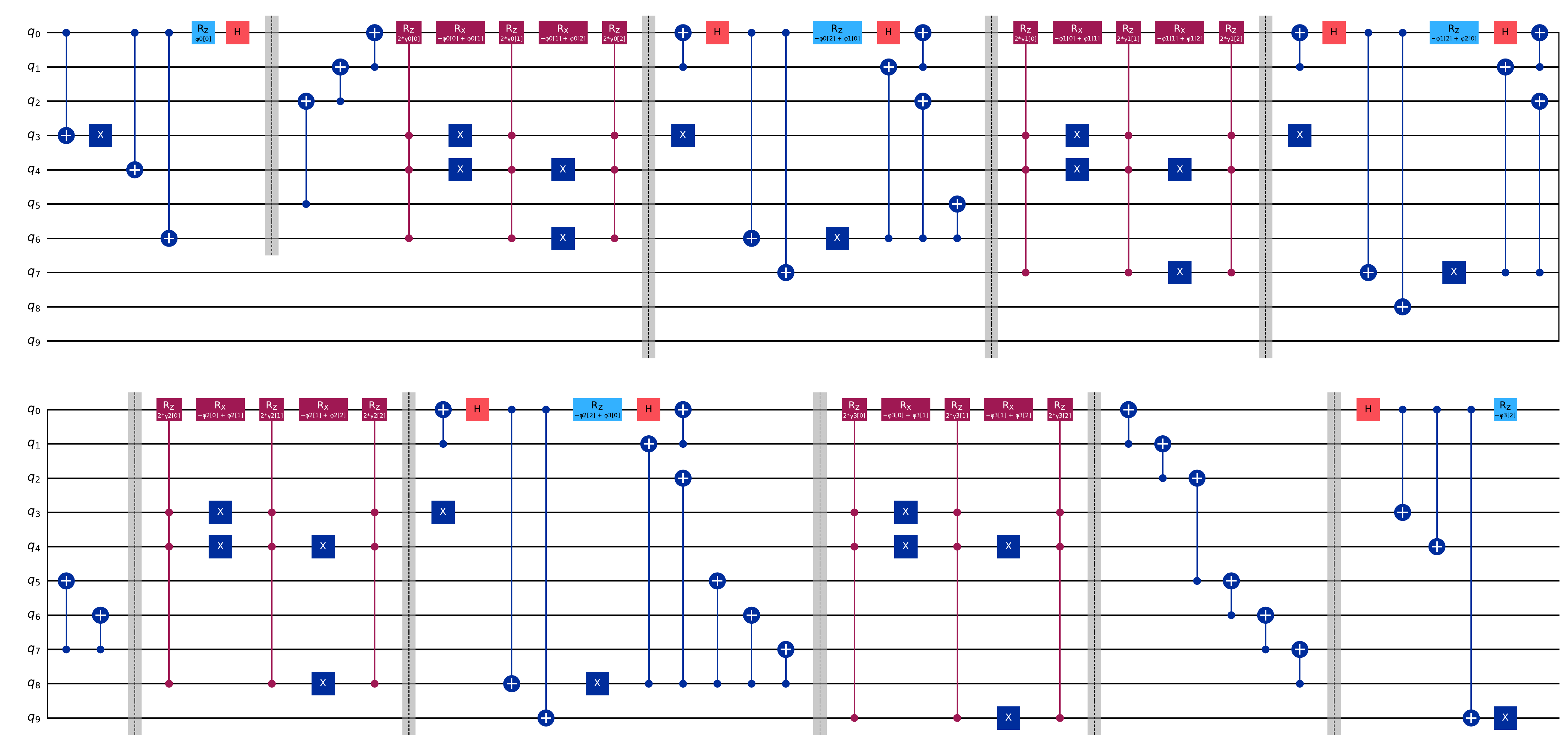}} \\
    \subfloat[After applying the pattern transform in Fig.~\ref{fig:CX_pat_b}\label{fig:fleet_opt_b}]{\includegraphics[width=0.98\linewidth]{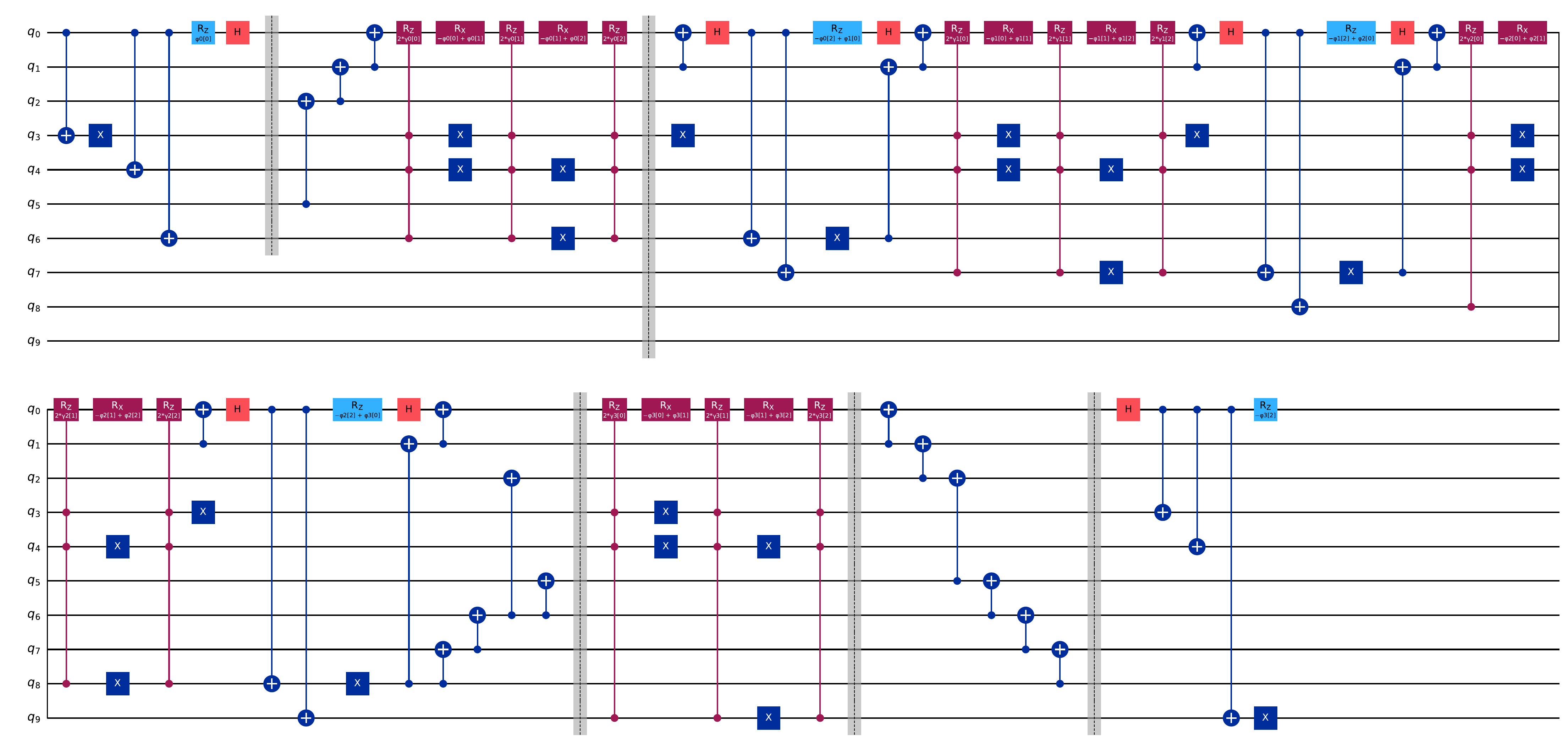}}
  \caption{
Quantum circuits during a cascade of CX reductions for the time evolution circuit of four triads.%
The barriers are inserted for better visibility.
(a) After applying the pattern transform in Fig.~\ref{fig:CX_pat_a}
in the boundary zones $B_6$, $B_7$, and $B_8$.
(b) After applying the pattern transform in Fig.~\ref{fig:CX_pat_b}
across the boundary zones for $B_7$ and $B_8$ and then for $B_6$ and (updated) $B_8$.
}
\end{figure*}

We are now ready to explain a cascade of CX reductions in and across the boundary zones,
using the time evolution circuit for four triads in Fig.~7(a) as an example.
First,
we see that the CX gates in the parity encoding sub-circuits in $B_k~(k=6,7,8)$ form a V-staircase pattern.  
We thus apply the pattern transform in Fig.~\ref{fig:CX_pat_a}, giving the circuit in Fig.~\ref{fig:fleet_opt_a}.
Note that two $CX_{1,0}$ gates must be left unchanged 
because of the intervening gates.
Next,
we apply the pattern transform in Fig.~\ref{fig:CX_pat_b} for $B_7$ and $B_8$ and then for $B_6$ and (updated) $B_8$,
yielding the circuit in Fig.~\ref{fig:fleet_opt_b}.
Note that the two $CX_{1,0}$ and $CX_{k,0}$ gates for $B_k$ are not movable due to dependence relations.
Finally, 
we apply the pattern transform in Fig.~\ref{fig:CX_pat_c} to $B_8$ and 
$PE_0^{\mathcal{I}(0,3) \cup \mathcal{I}(4,9)}$ (a descending staircase pattern). 
Note that two $CX_{1,0}$ are intact.
We have now obtained the optimized circuit in Fig.~\ref{fig:fleet_opt}.

We note that there are numerous optimization paths
from Fig.~\ref{fig:fleet}
to Fig.~\ref{fig:fleet_opt}.
Above, we described the optimization path
that is easiest for a human to follow.
In fact, our circuit optimizer takes a very different path.

Regarding the implementation of the circuit optimizer,
we extended the Qiskit transpiler \cite{qiskit2024}
to enable the above-mentioned cascade of CX reductions.
In particular,
we enhanced its template-matching capabilities
by adding new templates for equivalence relations
not supported by default \cite{iten2022template}.
We note that an extensible and customizable tool is crucial for implementing research prototypes like ours.

\section{Decomposition of the three C3RZ gates in the real-valued triad}
\label{app:threeC3RZs}

\begin{figure}[htbp]
\includegraphics[width=0.36\textwidth]{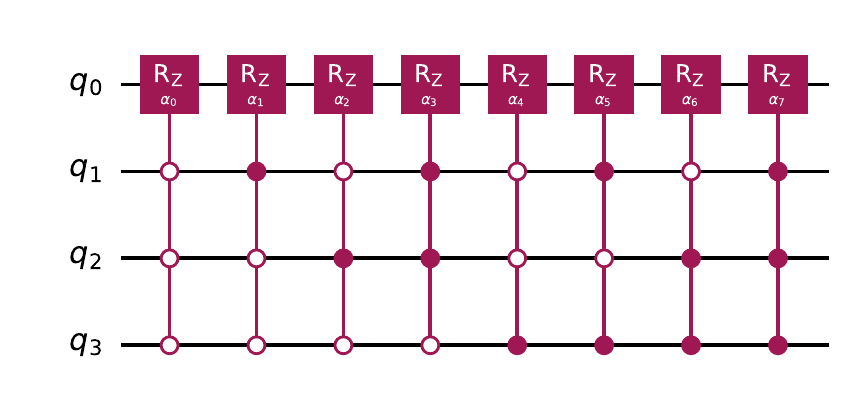}
\caption{
Uniformly controlled rotation
with three control qubits.
For a computational basis state
$\lvert q_3 q_2 q_1 \rangle = \lvert k \rangle$,
exactly $RZ(\alpha_k)$ is applied to $q_0$.
}
\label{fig:ucrot}
\end{figure}

Our decomposition simply follows the method in Ref.~\cite{Mottonen04}
for realizing a uniformly controlled rotation.
Fig.~\ref{fig:ucrot} shows such a rotation with three control qubits,
$q_1$, $q_2$, and $q_3$.
The indices for the rotation angles are chosen such that,
for a computational basis state
$\lvert q_3 q_2 q_1 \rangle = \lvert k \rangle$,
exactly $RZ(\alpha_k)$ is applied to $q_0$.
For our specific case in Fig.~\ref{fig:triad6430_r_opt},
$\alpha_6=r[0]$,
$\alpha_5=r[1]$,
$\alpha_3=r[2]$, and 
$\alpha_k=0$ otherwise. 
Note that black control circles bracketed by X gates in our figures are interpreted as white circles here.

\begin{figure}[htbp]
\includegraphics[width=0.48\textwidth]{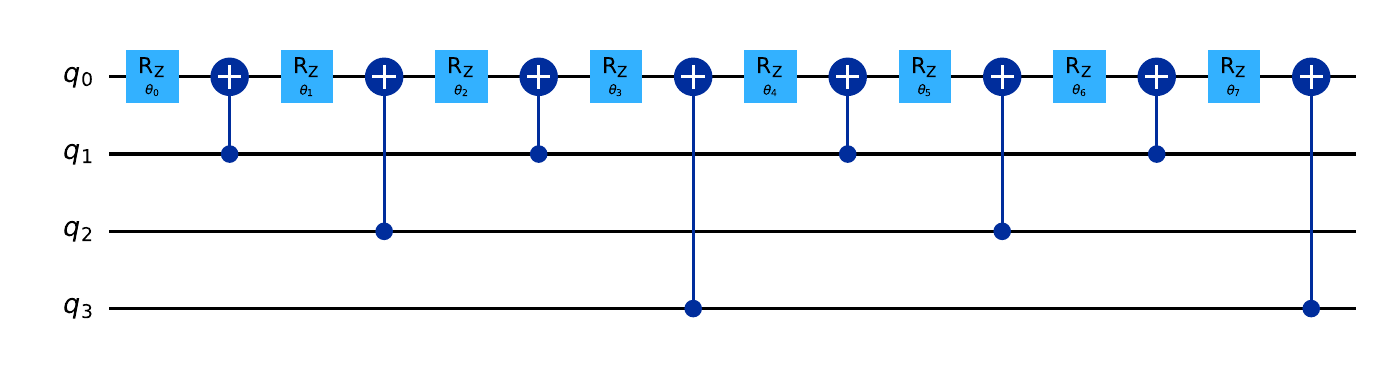}
\caption{
Decomposition
of a uniformly controlled rotation
in Fig.~\ref{fig:ucrot}
based on the binary reflected Gray code.
The rotation angles 
$\theta_j$ ($j=0,\ldots,7$)
can be obtained from $\alpha_j$ by solving a linear system.
}
\label{fig:ucrot_decomp}
\end{figure}

They showed that a uniformly controlled rotation
with $n$ control qubits
can be decomposed with $2^n$ CX gates based on the binary reflected gray code.  Fig.~\ref{fig:ucrot_decomp} presents such a decomposition
for Fig.~\ref{fig:ucrot} using the gray code
\begin{align}
g=[000, 001, 011, 010, 110, 111, 101, 100].
\nonumber
\end{align}
The $l$-th CX is inserted with $q_0$ (target) and $q_{m+1}$ (control),
when $g[l-1]$ and $g[l]$ differs in bit~$m$. 
When the circuit is executed
for a given computational basis state 
$\lvert q_3 q_2 q_1 \rangle = \lvert k \rangle$, 
$\theta_j$ is negated
when the Hamming weight of the bitwise-and of $k$ and $g[j]$ is odd.

Given $\alpha_j$ for $j=0,\ldots,7$, we can obtain
$\theta_j$ by solving the following linear system:
\begin{align}
\begin{pmatrix}
 1 &  1 &  1 &  1 &  1 &  1 &  1 &  1 \\
 1 & -1 & -1 &  1 &  1 & -1 & -1 &  1 \\
 1 &  1 & -1 & -1 & -1 & -1 &  1 &  1 \\
 1 & -1 &  1 & -1 & -1 &  1 & -1 &  1 \\
 1 &  1 &  1 &  1 & -1 & -1 & -1 & -1 \\
 1 & -1 & -1 &  1 & -1 &  1 &  1 & -1 \\
 1 &  1 & -1 & -1 &  1 &  1 & -1 & -1 \\
 1 & -1 &  1 & -1 &  1 & -1 &  1 & -1
\end{pmatrix}
\begin{pmatrix}
\theta_0 \\
\theta_1 \\
\theta_2 \\
\theta_3 \\
\theta_4 \\
\theta_5 \\
\theta_6 \\
\theta_7
\end{pmatrix} =
\begin{pmatrix}
\alpha_0 \\
\alpha_1 \\
\alpha_2 \\
\alpha_3 \\
\alpha_4 \\
\alpha_5 \\
\alpha_6 \\
\alpha_7
\end{pmatrix}
\nonumber
\end{align}

Let $M$ denote the matrix.
Since $\frac{1}{2\sqrt{2}}M$ is orthogonal,
$M^{-1} = \frac{1}{8}M^{T}$. Multiplying $M^{-1}$ by the $\alpha$ vector gives the solution.

\bibliographystyle{unsrt}
\bibliography{reference}

\end{document}

%% file: lsd_abstract_ver3to.tex
Electronic Hamiltonian simulation is commonly formulated by mapping fermionic operators to qubit operators and subsequently expanding the resulting ladder-operator products into Pauli strings. 
While general, this procedure obscures the higher-level fermionic structure and can hide opportunities for circuit optimization. Building on ladder-string-pair (Lasp) diagonalization originally developed for Hamiltonian simulation of partial differential equations, we construct time-evolution circuits for the second-quantized electronic Hamiltonian without performing a Pauli expansion. By combining the Jordan--Wigner transformation with Lasp diagonalization, we obtain compact circuits composed of GHZ-state transformations, parity-encoding circuits, and multi-controlled $Z$ rotations.
For the most general case of complex-valued coefficients, we present a time-evolution circuit for a two-body fermionic Lasp operator that corresponds to 16 Pauli strings in the Pauli-expansion approach but does not suffer from Trotter error at this stage. 
Expanding the optimization scope from a single operator to a triad of three fermionic Lasp operators sharing the same four spin-orbital indices enables systematic cancellation of CX gates,
reducing the CX-gate count from 36 to 12 in the example considered, still without introducing Trotter error at this stage. 
For $n$ spin orbitals, further expanding the optimization scope to a sequence of $O(n)$ suitably ordered triads enables a cascade of CX-gate reductions across triad boundaries, reducing the CX-gate count from $O(n^2)$  to $O(n)$. 
We also show that, in the Lasp-based formulation, a controlled version of a time-evolution circuit can be obtained simply by adding an extra control to its multi-controlled $Z$ rotations.
Finally, while a triad circuit in the complex-valued case contains three C3RZ ($Z$ rotation controlled by three qubits) gates, the corresponding real-valued circuit contains the same three gates but in a back-to-back configuration, 
forming a uniformly controlled rotation that can be decomposed using only eight CX gates in total.
These results demonstrate that the Lasp-based approach enables more efficient time-evolution circuits by preserving high-level circuit structures and thereby expanding the scope of optimization, providing a systematic route toward more efficient electronic Hamiltonian simulation.

%% file: lsd_introduction_ver3to.tex
Hamiltonian simulation is one of the central primitives in quantum computing,
underlying algorithms for quantum dynamics, phase and energy estimation, and
many other applications~\cite{berry2015simulating,bosse2025efficient,lin2022heisenberg}.
Electronic-structure simulation is a particularly important application of
quantum computing~\cite{alexeev2025perspective}, where the second-quantized
electronic Hamiltonian is mapped to a qubit Hamiltonian and its time-evolution
operator is implemented as a quantum circuit. A standard route is to transform
fermionic creation and annihilation operators into qubit operators, for example
by the Jordan--Wigner transformation~\cite{JordanWigner1928,Seeley2012}, and
subsequently expand the resulting products of ladder operators into Pauli
strings.

The Pauli-string representation is general and convenient, and a variety of
methods have been developed to optimize Hamiltonian-simulation circuits at the
level of individual Pauli terms or structured groups of Pauli
terms~\cite{van2020circuit,li2022paulihedral,mukhopadhyay2023synthesizing,
decker2025kernpiler}. However,
expanding a fermionic Hamiltonian into Pauli strings
at an early stage can obscure the higher-level structure inherited
from the original fermionic operators. Relations among terms that are
transparent in the fermionic representation may consequently become difficult
to exploit during circuit construction and optimization. This motivates a
complementary strategy in which the ladder-operator structure is retained for
as long as possible.

A direct precursor of the present work is the Hamiltonian-simulation algorithm
for linear hyperbolic partial differential equations proposed by
Sato et al.~\cite{sato2024hamiltonian}. In that setting, finite-difference
operators give rise naturally to tensor products of ladder operators. A
straightforward Pauli expansion can generate an exponentially large number of
Pauli terms, whereas Sato et al. showed that conjugate pairs of ladder strings
can instead be diagonalized directly,
enabling scalable circuit implementations using multi-controlled rotations.
This Lasp (Ladder string pair)-based construction was
subsequently applied to more general PDE
settings~\cite{hul2024quantum,sato2025quantum}
and opened up new possibilities for significant optimization of PDE circuits~\cite{MGSO2026}.
The present work extends this
Lasp-based approach from PDE Hamiltonians to the second-quantized
electronic Hamiltonian.

The electronic Hamiltonian introduces structures not encountered in the
previous PDE applications. In particular, the Jordan--Wigner transformation
produces parity strings between fermionic ladder operators, while the two-body
Hamiltonian contains related excitation terms associated with common sets of
spin-orbital indices. We show that these features can be accommodated naturally
within the Lasp-based framework without performing a Pauli expansion:
Jordan--Wigner strings are incorporated through parity encoding, while
pairs of fermionic two-body terms retain a compact Lasp-based
representation at the circuit-construction level.

In the present work, we show that this representation exposes a natural
higher-level organization of the two-body Hamiltonian. For four distinct
spin-orbital indices, three related fermionic Lasp operators form a commuting
triad that can be treated as a single optimization unit. This collective
treatment reveals entangling-gate cancellations that are much less transparent
after Pauli expansion and provides the basic building block for our circuit construction and
optimization.

Furthermore, we show that the same principle extends beyond an individual
triad. Suitably ordering sequences of triads enables cancellations across
triad boundaries and, for the families analyzed here, reduces the
CX-gate count scaling from quadratic to linear in the number of spin orbitals.
Additional simplifications are obtained for real-valued Hamiltonians and for
controlled time evolution. Our general Lasp formulation also includes the
double-qubit excitation circuit of Yordanov et al.~\cite{Yordanov20} as a
special case.

The remainder of this paper is organized as follows.
Section~\ref{sec:lasp} introduces Lasp diagonalization and the
corresponding time-evolution circuits.
Section~\ref{sec:const_circ} applies the Lasp-based construction to the
second-quantized electronic Hamiltonian and derives circuits for its one- and two-body terms. 
We then turn to novel circuit optimizations enabled by our Lasp formulation.
Section~\ref{sec:triad} presents optimization of a triad,
Section~\ref{sec:fleet} discusses optimization of suitably ordered triads,
Section~\ref{sec:cH} considers controlled versions of time-evolution circuits, 
and Section~\ref{sec:real} presents further optimization for real-valued Hamiltonians. 
Finally,
Section~\ref{sec:summary} concludes the paper.

%% file: lsd_summary_ver3to.tex
We presented a Lasp-based construction of time-evolution circuits for the second-quantized electronic Hamiltonian. Instead of expanding fermionic operators into Pauli strings, we diagonalize conjugate pairs of fermionic operators. Combined with the Jordan--Wigner transformation, this direct diagonalization yields compact time-evolution circuits composed of GHZ-state-preparation subcircuits, parity-encoding subcircuits, and controlled $Z$ rotations.
We treated both the one- and two-body parts of the electronic Hamiltonian, considering all index configurations. We primarily focused on the most general case of complex-valued coefficients and obtained compact time-evolution circuits for fermionic one-body and two-body Lasp operators, which correspond to four and 16 Pauli strings, respectively, in the Pauli-expansion approach, but do not suffer from Trotter errors at this stage. This illustrates an important advantage of retaining higher-level circuit representations rather than immediately decomposing the Hamiltonian into Pauli strings.

We then expanded the optimization scope from a single operator to a triad. For the two-body Hamiltonian with four distinct spin-orbital indices, three fermionic Lasp operators sharing the same four indices form a triad. Treating the triad as an optimization unit enables systematic cancellation of entangling gates; in the specific example considered in this work, the CX-gate count is reduced from 36 to 12 while the three C3RZ gates are retained. In addition, because the operators within the triad mutually commute, the resulting circuit still does not suffer from Trotter errors at this stage.

We further expanded the optimization scope from a triad to a collection of triads. For suitably ordered $O(n)$ triads, a cascade of CX-gate reductions occurs across triad boundaries, reducing the CX-gate count from $O(n^2)$ to $O(n)$. We also considered what constitutes a suitable ordering: the triads should share a fixed rotation qubit, while adjacent triads should differ in only one qubit index, with the index displacement kept as small as possible. However, a systematic search for such orderings, as well as for efficient compositions of multiple groups of ordered triads, 
remains an important direction for future work.

We also considered controlled versions of time-evolution circuits, showing that, in the Lasp-based formulation, a controlled version can be obtained simply by adding an extra control to the multi-controlled $Z$ rotations.
Finally, we considered the real-valued case, in which the time-evolution circuit contains three back-to-back C3RZ gates with mutually exclusive control conditions, forming an instance of a uniformly controlled rotation. This structure allows a highly efficient decomposition: the three C3RZ gates can be collectively decomposed using only eight CX gates in total.

Overall, these results demonstrate that the Lasp-based approach enables more efficient time-evolution circuits by preserving high-level circuit structures and thereby expanding the scope of optimization. More broadly, our work highlights the potential of high-level circuit construction and optimization as a systematic route toward more efficient Hamiltonian simulation.

%% file: lsd_acknowledgment_sato.tex
This work was also supported in part by JSPS KAKENHI
(Grant Nos. JP25K01688),
JST COI-NEXT (Grant No. JPMJPF2221),
MEXT Q-LEAP (Grant No. JPMXS0118067246),
IBM-UTokyo lab, 
and the RIKEN TRIP initiative.